\documentclass[twoside]{article}
\usepackage{babel}
\usepackage{epsfig}
\usepackage{amsmath}
\usepackage{amssymb}
\usepackage{amsfonts}
\usepackage{pifont}

\catcode`\@=11
\long\def\@makefntext#1{
\protect\noindent \hbox to 3.2pt {\hskip-.9pt  
$^{{\eightrm\@thefnmark}}$\hfil}#1\hfill}               %CAN BE USED 

\def\@makefnmark{\hbox to 0pt{$^{\@thefnmark}$\hss}}    %ORIGINAL 
        
\def\ps@myheadings{\let\@mkboth\@gobbletwo
\def\@oddhead{\hbox{}
\rightmark\hfil\eightrm\thepage}   
\def\@oddfoot{}\def\@evenhead{\eightrm\thepage\hfil
\leftmark\hbox{}}\def\@evenfoot{}
\def\sectionmark##1{}\def\subsectionmark##1{}}

\oddsidemargin=\evensidemargin
\newcounter{sectionc}\newcounter{subsectionc}\newcounter{subsubsectionc}
\renewcommand{\section}[1] {\vspace{12pt}\addtocounter{sectionc}{1} 
\setcounter{subsectionc}{0}\setcounter{subsubsectionc}{0}\noindent 
\protected@edef\@currentlabel{\thesectionc}
        {\tenbf\thesectionc. #1}\par\vspace{5pt}}
\renewcommand{\subsection}[1] {\vspace{12pt}\addtocounter{subsectionc}{1} 
        \setcounter{subsubsectionc}{0}\noindent 
         \protected@edef\@currentlabel{\thesectionc.\thesubsectionc}
        {\bf\thesectionc.\thesubsectionc. {\kern1pt \bfit #1}}\par\vspace{5pt}}
\renewcommand{\subsubsection}[1] {\vspace{12pt}\addtocounter{subsubsectionc}{1}
\protected@edef\@currentlabel{\thesectionc.\thesubsectionc.\thesubsubsectionc}
        \noindent{\tenrm\thesectionc.\thesubsectionc.\thesubsubsectionc.
        {\kern1pt \tenit #1}}\par\vspace{5pt}}
\newcommand{\nonumsection}[1] {\vspace{12pt}\noindent{\tenbf #1}
        \par\vspace{5pt}}

\newcounter{appendixc}
\newcounter{subappendixc}[appendixc]
\newcounter{subsubappendixc}[subappendixc]
\renewcommand{\thesubappendixc}{\Alph{appendixc}.\arabic{subappendixc}}
\renewcommand{\thesubsubappendixc}
        {\Alph{appendixc}.\arabic{subappendixc}.\arabic{subsubappendixc}}

\renewcommand{\appendix}[1] {\vspace{12pt}
        \refstepcounter{appendixc}
        \setcounter{figure}{0}
        \setcounter{table}{0}
        \setcounter{lemma}{0}
        \setcounter{theorem}{0}
        \setcounter{corollary}{0}
        \setcounter{definition}{0}
        \setcounter{equation}{0}
        \renewcommand{\thefigure}{\Alph{appendixc}.\arabic{figure}}
        \renewcommand{\thetable}{\Alph{appendixc}.\arabic{table}}
        \renewcommand{\theappendixc}{\Alph{appendixc}}
        \renewcommand{\thelemma}{\Alph{appendixc}.\arabic{lemma}}
        \renewcommand{\thetheorem}{\Alph{appendixc}.\arabic{theorem}}
        \renewcommand{\thedefinition}{\Alph{appendixc}.\arabic{definition}}
        \renewcommand{\thecorollary}{\Alph{appendixc}.\arabic{corollary}}
        \renewcommand{\theequation}{\Alph{appendixc}.\arabic{equation}}
        \noindent{\tenbf Appendix \theappendixc #1}\par\vspace{5pt}}
\newcommand{\subappendix}[1] {\vspace{12pt}
        \refstepcounter{subappendixc}
        \noindent{\bf Appendix \thesubappendixc. {\kern1pt \bfit #1}}
        \par\vspace{5pt}}
\newcommand{\subsubappendix}[1] {\vspace{12pt}
        \refstepcounter{subsubappendixc}
        \noindent{\rm Appendix \thesubsubappendixc. {\kern1pt \tenit #1}}
        \par\vspace{5pt}}

\newcommand{\textlineskip}{\baselineskip=13pt}
\newcommand{\smalllineskip}{\baselineskip=10pt}

\def\eightcirc{
\begin{picture}(0,0)
\put(4.4,1.8){\circle{6.5}}
\end{picture}}
\def\eightcopyright{\eightcirc\kern2.7pt\hbox{\eightrm c}}

\def\abstracts#1#2#3{{
        \centering{\begin{minipage}{4.5in}\baselineskip=10pt\footnotesize
        \parindent=0pt #1\par 
        \parindent=15pt #2\par
        \parindent=15pt #3
        \end{minipage}}\par}} 

\renewenvironment{thebibliography}[1]
        {\frenchspacing
         \ninerm\baselineskip=11pt
         \begin{list}{\arabic{enumi}.}
        {\usecounter{enumi}\setlength{\parsep}{0pt}
         \setlength{\leftmargin 12.7pt}{\rightmargin 0pt} %FOR 1--9 ITEMS
         \setlength{\itemsep}{0pt} \settowidth
        {\labelwidth}{#1.}\sloppy}}{\end{list}}

\newcounter{itemlistc}
\newcounter{romanlistc}
\newcounter{alphlistc}
\newcounter{arabiclistc}

\newcommand{\fcaption}[1]{
        \refstepcounter{figure}
        \setbox\@tempboxa = \hbox{\footnotesize Fig.~\thefigure. #1}
        \ifdim \wd\@tempboxa > 5in
           {\begin{center}
        \parbox{5in}{\footnotesize\smalllineskip Fig.~\thefigure. #1}
            \end{center}}
        \else
             {\begin{center}
             {\footnotesize Fig.~\thefigure. #1}
              \end{center}}
        \fi}

\newcommand{\tcaption}[1]{
        \refstepcounter{table}
        \setbox\@tempboxa = \hbox{\footnotesize Table~\thetable. #1}
        \ifdim \wd\@tempboxa > 5in
           {\begin{center}
        \parbox{5in}{\footnotesize\smalllineskip Table~\thetable. #1}
            \end{center}}
        \else
             {\begin{center}
             {\footnotesize Table~\thetable. #1}
              \end{center}}
        \fi}

\def\@citex[#1]#2{\if@filesw\immediate\write\@auxout
        {\string\citation{#2}}\fi
\def\@citea{}\@cite{\@for\@citeb:=#2\do
        {\@citea\def\@citea{,}\@ifundefined
        {b@\@citeb}{{\bf ?}\@warning
        {Citation `\@citeb' on page \thepage \space undefined}}
        {\csname b@\@citeb\endcsname}}}{#1}}

\newif\if@cghi
\def\cite{\@cghitrue\@ifnextchar [{\@tempswatrue
        \@citex}{\@tempswafalse\@citex[]}}
\def\citelow{\@cghifalse\@ifnextchar [{\@tempswatrue
        \@citex}{\@tempswafalse\@citex[]}}
\def\@cite#1#2{{$\null^{#1}$\if@tempswa\typeout
        {IJCGA warning: optional citation argument 
        ignored: `#2'} \fi}}

\def\pmb#1{\setbox0=\hbox{#1}
        \kern-.025em\copy0\kern-\wd0
        \kern.05em\copy0\kern-\wd0
        \kern-.025em\raise.0433em\box0}

\def\fnt#1#2{\footnotetext{\kern-.3em
        {$^{\mbox{\scriptsize #1}}$}{#2}}}

\font\tenrm=cmr10
\font\tenit=cmti10 
\font\tenbf=cmbx10
\font\bfit=cmbxti10 at 10pt
\font\ninerm=cmr9

\font\eightrm=cmr8

\def\qed{\hbox{${\vcenter{\vbox{                        %HOLLOW SQUARE
   \hrule height 0.4pt\hbox{\vrule width 0.4pt height 6pt
   \kern5pt\vrule width 0.4pt}\hrule height 0.4pt}}}$}}

\newcommand{\E}[1]{10$^{#1}$~eV}
                          
\begin{document}
\newcommand{\bi}{\begin{itemize}}
\newcommand{\ei}{\end{itemize}}
\vspace*{0.88truein}

\centerline{\bf PHYSICS OF EXTREMELY HIGH ENERGY COSMIC RAYS}
\vspace*{0.37truein}
\centerline{\footnotesize XAVIER BERTOU, MURAT BORATAV, ANTOINE LETESSIER-SELVON
}
\vspace*{0.015truein}
\centerline{\footnotesize\it Laboratoire de Physique Nucl\'eaire et des Hautes
Energies (IN2P3/CNRS),}
\baselineskip=10pt
\centerline{\footnotesize\it Universit\'es Paris 6\&7, 4 place Jussieu 75005 Paris, France}
\vspace*{10pt}

\abstracts{Over the last third of the century, a few tens of events,
  detected by ground-based cosmic ray detectors, have opened a new
  window in the field of high-energy astrophysics. These events have
  macroscopic energies -~exceeding $5\times10^{19}$ eV~-, unobserved sources
  -~if supposed to be in our vicinity~-, an unknown chemical composition and
  a production and transport mechanism yet to be explained. 
  With a flux as low as one particle per century per square kilometer, only
  dedicated detectors with huge apertures can bring in the high-quality and
  statistically significant data needed to answer those questions. In
  this article, we review the present status of the field both 
  from an experimental and theoretical point of view.
  Special attention is given to the next generation of detectors devoted to the thorough 
  exploration of the highest energy ranges.}{}{}

\vspace*{1pt}\textlineskip  

\section{Introduction} 
\label{intro}
\noindent 
This review article will mainly concern the problems raised by the existence
and observation of cosmic rays whose energies are above
$5\times10^{19}$ eV. Such cosmic rays -~for which we shall use the term
``extremely  high energy cosmic rays" or EHECR~- are exceptional for the
following reasons:

\bi

\item The  Greisen-Zatsepin-Kuzmin (GZK) cutoff\cite{Greisen} corresponds to
the threshold for inelastic collisions between the cosmic microwave background
(CMB) and protons (photo-pion production) or heavy nuclei
(photo-disintegration). Similar cutoffs exist at lower energies for gammas interacting with
background photons (CMB, infra-red or radio waves). Consequently, and except for neutrinos, if the EHECR
observed on Earth are due to the known stable particles, they must be produced in our vicinity.
At the GZK cutoff, the ``visible" universe shrinks suddenly to a sphere of a few tens of 
megaparsecs (Mpc).\footnote{The megaparsec is the cosmological unit of distance that we
shall use throughout this article. $1\, \text{Mpc} =
3.26\times10^6\,\text{light-years} \approx 3\times10^{19}\, \text{km}$}~ This
should be reflected on the energy spectrum of cosmic rays as a sharp drop
around $5\times 10^{19}$ eV. The available data show no such drop. On the
contrary, if a structure exists in the energy spectrum, it is in the sense of a
softening of the power law, possibly suggesting a new component rising from
under an otherwise steeply falling spectrum.

\item There are very few conventional astrophysical sources
considered by the experts as being able to accelerate particles at
energies exceeding those of the most energetic EHECR that have been
observed in the past. 

\item At such energies, the bending effect of the galactic and extragalactic magnetic
fields are quite weak on the EHECR. Thus, even if they are charged particles,
their reconstructed incident direction should point toward their sources within
a few degrees. This distinguishes the EHECR from their counterparts in the lower
energy regions: one can use them for point-source-search astronomy. 

\ei

The widely shared excitement about the EHECR  
comes from the
above considerations and from the study of the scarce
data available. The EHECR exist. They have to be produced somewhere and with a
mechanism we do not understand up to now: to explain the observation of the
highest energy events, the energy at the source should probably be in the ZeV
range (Z is for \emph{zetta}, i.e. $10^{21}$), namely at least 150 joules! If
these extraordinary accelerating engines are astrophysical macroscopic objects,
they must be visible through some counterpart that e.g. optical- or radio-astronomy
should detect. But no such remarkable object is visible in the directions from
where the EHECR come. There is even no convincing evidence that one can find any
correlation between the incoming directions and the inhomogeneous distribution
of matter in our vicinity. 

During the 35 years where the EHECR puzzle was fed by a slow input of experimental data,
many models and theories were put forward to try to explain it.
In the following, we shall develop in detail the facts and arguments briefly
mentioned in this introduction. In section 2 we discuss the 
cosmic ray interactions in the atmosphere and the detection 
techniques. In section 3 we review our current knowledge of the subject while section 4 
is devoted to transport problems and to candidate source characteristics. Finally section 5 
describes the next generation of detectors devoted to EHECR studies.

There is no doubt that for the scientists working in the field of what is now called 
``Particle Astrophysics" the near
future will be a thrilling period where a large variety of models or theories 
will be confronted with high-quality and high-statistical data brought by a new 
generation of detectors.

To avoid repetitive use of large powers of ten, the energy units in the
following will mostly be in zetta-electron-volts (ZeV, see above) and
exa-electron-volts (EeV, i.e. $10^{18}$ eV). 

\section{The detection of the EHECR}
\noindent
When the cosmic ray flux becomes smaller than 1 particle per m$^2$ 
per year, satellite borne detectors are not appropriate any more. This
happens above $10^{16}$ eV (the so-called ``knee" region). Then 
large surfaces are needed. The detectors then become ground-based. What they
detect is not the incident particle itself but the cascade of
secondary particles initiated by the cosmic ray interactions
with the molecules of the atmosphere. The object of
study is therefore this cascade or shower called the \emph{Extensive Air Shower}
(EAS). 

Another distinction has to be made between the techniques used at lower and
higher energies. Air showers can actually be observed at much lower energies than
the knee region. Gamma ray astronomy, for example, uses the properties of the EAS (production
of Cerenkov light by charged secondaries in the atmosphere) to detect and
measure cosmic photons from a few tens of GeV up. The main difference with the
higher energies is that above a few hundreds of TeV and up to \E{20}, cosmic rays are mainly
charged particles which cannot be associated with a point source as is the
case with photons. Therefore a ground-based cosmic ray detector must survey the
whole sky and not point at a defined source.

There are two major techniques used in the detection of the highest energy
cosmic rays. The first, and the most frequent, is to build an array of sensors
(scintillators, water Cerenkov tanks, muon detectors) spread over a large area.
The detectors count the particle densities at any given moment, thus sampling
the EAS particles hitting the ground. The surface of the array 
is chosen in adequation with the incident flux and 
the energy range one wants to explore (large array surfaces for weak fluxes,
large density of sensors for lower energies). From this sampling of the 
lateral development of the shower at a given
atmospheric depth one can deduce the direction, the energy and possibly the identity 
of the primary CR. 
The second technique, until recently the exclusivity of a group from the University of
Utah, consists in studying the longitudinal development of the EAS by detecting
the fluorescence light produced by the interactions of the the charged secondaries.

\subsection{The Extensive Air Showers\label{ident}}
\noindent
On an incident cosmic ray the atmosphere acts as a calorimeter with
variable density, a vertical thickness of 26 radiation lengths and
about 11 interaction lengths. In the following, we will describe the
properties of a vertical EAS initiated by a 10~EeV proton and mention how
some of these properties are modified with energy and with the nature of
the initial cosmic ray (CR).

At sea level (atmospheric thickness of 1033~g/cm$^2$) the number of
secondaries reaching ground level (with energies in excess of 200~keV) is about $3\times 10^{10}$
particles. 99\% of these are photons and electrons/positrons in a ratio
of 6 to 1. Their energy is mostly in the range 1 to 10~MeV and they
transport 85\% of the total energy. The remaining 1\% is shared between
mostly muons with an average energy of 1~GeV (and carrying about 10\% of
the total energy), pions of a few GeV (about 4\% of the total energy)
and, in smaller proportions, neutrinos and baryons. The
lateral development of the shower is represented by its Moli\`ere radius
(or the distance within which 90\% of the total energy of the shower is
contained) which, in the standard air is 70~m. However, the actual
extension of the shower at ground level is of course much larger. As an
example, at a distance of 1~km from the shower axis, the average
densities of photons/electrons/muons are 30/2/1 per m$^2$. 

The longitudinal development of the EAS will be described to some extent
in the next section. Let us just mention that the maximum size of the
shower is reached at an atmospheric depth of 830 g/cm$^2$ (or an altitude
of about 1800 meters) and contains about $7\times 10^9$ electrons (which
produce the fluorescence light detected with the Fly's Eye telescopes,
see below).

Showers initiated by heavier nuclei can be described by making use of a 
superposition principle: a heavy nucleus of mass number $A$ and energy
$E$ can be in a first approximation considered as a superposition of $A$
showers initiated by nucleons each with an energy of $E/A$, therefore
less penetrating than a nucleon with energy $E$ (roughly 100 g/cm$^2$
higher in the atmosphere for iron). 

A ground array makes use of two main effects to separate heavy from light nuclei (and from photons): 
the proportion of muons compared to the electromagnetic component of the shower and the rise time of the detected
signal. Both parameters are due to the way the muons are produced during the shower development. The muons in a shower come from
the decay of charged pions when they reach an energy low enough so that
their decay length becomes smaller than their interaction length. Since
this happens earlier in the case of a primary heavy nucleus, the resulting shower is richer in muons than 
 a proton shower. At the same time, and since muons are produced
earlier in the shower development, they reach the ground also earlier,
compared to the electromagnetic component which undergoes many
interactions before reaching the detector array. 

For a photon shower the proportion of muons will be even smaller and 
at the highest energies and another physical process will have important
consequences on the EAS detection and characterization. This is the
Landau-Pomeranchuk-Migdal (LPM) effect\cite{LPM} which describes the
decrease of the photon/electron nucleus cross-sections with energy
and with the density of the medium with which they
interact. 
Even in the upper atmosphere, the LPM effect becomes appreciable at energies in the EeV range so 
that it is possible for a photon of 100~EeV to develop an
EAS very deep in the atmosphere, yielding less than $10^9$ particles at ground level.
Such a shower would have an extension of only a few km$^2$. 

These effects are studied through heavy use of EAS Monte Carlo programs
such as AIRES,\cite{Aires} CORSIKA,\cite{Corsika} HEMAS\cite{Hemas} or
MOCCA.\cite{Mocca}
At the EHECR ranges, where the center-of-mass energies are much higher (almost two orders of magnitude) than those attainable in the future (and the most powerful) accelerator LHC, 
the correct modelling of the EAS in these programs becomes delicate. 
Some
data are available from accelerator experiments such as HERA,\cite{Hera}
and showers of about $10^{16}$~eV are now being well studied through
experiments such as KASCADE.\cite{Kascade} The models are thus constrained at
lower energies and then extrapolated at higher ranges.

The most commonly used models for the high energy hadronic interactions of the simulation programs are
SIBYLL,\cite{Sibyll} VENUS,\cite{Venus} QGSJet\cite{QGSJet} and
DPMJet.\cite{DPMJet}  Interactions at lower energies are either processed
through internal routines of the EAS simulation programs or by well-known packages
such as GHEISHA.\cite{Gheisha}  Some detailed studies of the different models
are available.\cite{Comparison} A clear conclusion is that  the simulation
results are never identical, even when the same theoretical models are used in
different programs.  However, when simulating a shower, these models are
only used for the first few interactions and  an
EAS yields about $10^{10}$ or $10^{11}$ particles at ground level. Therefore the main
shower parameters, such as the reconstructed direction and energy of the
primary CR, are never strongly dependent on the chosen model. However,
the identification of the primary is more problematic.
Whatever technique is chosen (see Section 3.4 for details) the
parameters used to identify the primary cosmic ray undergo large physical
fluctuations which make an unambiguous identification difficult.

A complete analysis done by the KASCADE group on the hadronic core of
EAS\cite{Kascade} has put some constraints on interaction models beyond accelerator energies.
Various studies seem to indicate QGSJet as being the model which best
reproduces the data\cite{Comparison,Erlykin} with still some disagreement
at the {\em knee} energies ($10^{16}$~eV). For the highest energies,
additionnal work (and data) is needed to improve the agreement between the available models.

\subsection{The optical fluorescence technique}
\noindent
The idea that one could use the fluorescence light produced in the atmosphere
to detect and characterize the EAS was first suggested independently by Greisen and Suga\cite{greisen}
and then by a few other authors in the early sixties. The basic principle is
simple\cite{Cassiday} (although the detector itself and the measuring
techniques are quite sophisticated): the charged particles produced in the
development of the EAS excite the nitrogen atoms of the atmosphere which then
emit, very quickly and isotropically, fluorescence light which can be detected by a
photo-multiplier. The emission efficiency (ratio of the energy emitted as fluorescence light to the deposited one)
is poor (less than 1\%), therefore observations can only be done on clear
moonless nights (which results in an average 10\% duty cycle) and low
energy showers can hardly be observed. At higher
energies, the huge number of particles in the shower\footnote{The highest
energy shower ever detected (320~EeV) was observed by 
the Fly's Eye detector: at the shower maximum, the number of particles was larger than
$2\times10^{11}$.}~ produce enough light to be detected even at large distances. The
fluorescence yield is 4 photons per electron per meter at ground level. The emitted
light is typically in the 300-400 nm UV range to which the atmosphere is quite
transparent. Under favorable atmospheric conditions an EHECR shower
can be detected at distances as large as
20 km (about two attenuation lengths in a standard desert atmosphere 
at ground level). 

The first successful detectors based on these ideas were built by a group of the University of Utah, 
under the  name of ``Fly's Eyes", and used with the Volcano Ranch ground array 
(see Section~\ref{vr}). A complete detector
was then installed at Dugway (Utah) and started to take data in 1982. An updated
version, the High-Resolution Fly's Eye, or HiRes, is presently running on this same site. 

\begin{figure}[!htb] 
 
\begin{center}   
\mbox{\epsfig{file=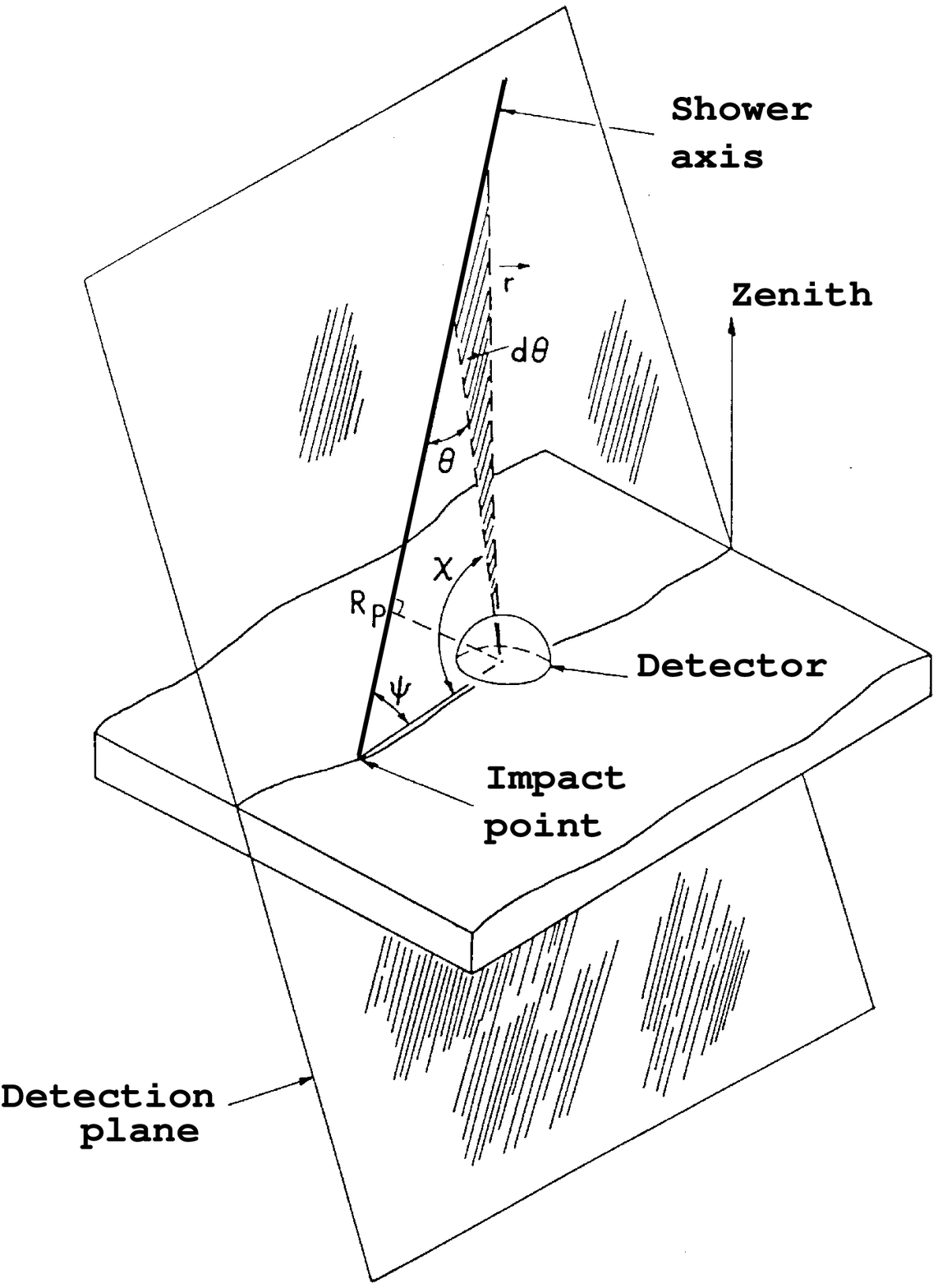,width=6cm}} 
\end{center}   

\fcaption{The principle of the detection of an EAS by a fluorescence 
telescope.\cite{Cassiday}\label{fe}} 
\end{figure} 
Figure \ref{fe} shows the geometry of the detection of an air shower by Fly's
Eye type detectors (which are usually given the  more generic name of ``fluorescence detectors'' or ``fluorescence telescopes''). The detector sees the shower as a variable light bulb\footnote{A rough estimate of the equivalent radiated power would be $3E_{18}$ watts at the shower maximum, where $E_{18}$ is the primary energy in EeV.}~ moving at
the speed of light along the shower axis. The detector itself is a set of
phototubes mounted on a ``camera" set at the focal plane of a mirror. Each
phototube sees a small portion of the sky (typically $1^{\circ}$). A fit on the
pattern of tubes hit by the fluorescence photons determines with a 
precision better than one degree the plane containing the detector and the
shower axis. In the \emph{stereo mode} (EAS seen by two telescopes installed a few km apart), two planes are thus reconstructed and their intersection gives the incident direction with good precision. In the \emph{mono mode} (EAS seen by a single telescope), one has to rely  on the the time of arrival of the photons on the tubes. A good reconstruction of the direction (the $\Psi$ angle) then needs a large number of pixels to be hit, enough to measure simultaneously the angular velocity and the angular acceleration of the shower development. Finally, in the \emph{hybrid mode}, i.e. simultaneous detection of the EAS with a fluorescence telescope \emph{and} a ground array, the knowledge of the intersection of the shower axis with the array plane (reconstructed by the array) allows the selection of the right direction in the family of lines in the detector-shower axis plane. For 100~EeV showers, a precision of $0.2^{\circ}$ can then be reached.

The fluorescence technique is the most appropriate way to measure the
energy of the incident cosmic ray: it is a partial calorimetric measurement with continuous longitudinal sampling. The amount
of fluorescence light emitted is proportional to the number of charged particles in the
shower. The EAS has a longitudinal development usually parametrized by the
analytic Gaisser-Hillas function giving the size $N_{\text{e}}$ of the shower
(actually the number of the ionizing electrons) as a function of atmospheric
depth $x$: 
$$
N_{\text{e}}(x)=N_{\text{max}}\left(\frac{x-x_0}{X_{\text{max}}-x_0}
\right)^{(X_{\text{max}}-x_0)/\lambda}\text{e}^{(X_{\text{max}}-x)/\lambda}
$$
where $\lambda=70\,$g/cm$^2$, $x_0$ is the depth at which the first interaction
occurs, and $X_{\text{max}}$ the position of the shower maximum. The total
energy of the shower is proportional to the integral of this function,
knowing that the average energy loss per particle is 2.2~MeV/g~cm$^{-2}$.

In practice several effects have to be taken into account to properly convert the detected
fluorescence signal into the primary CR energy. These include the subtraction of the
direct or diffused Cerenkov light, the (wavelength dependent) Rayleigh and Mie
(aerosol) scatterings, the dependence of the attenuation on altitude (and
elevation for a given altitude) and atmospheric conditions, the energy
transported by the neutral particles (neutrinos), the hadrons interacting with nuclei (whose energy is not
converted into fluorescence) and penetrating muons whose energy is mostly
dumped into the earth. One also has to take into account that a shower is never seen in its totality by a fluorescence telescope: the Gaisser-Hillas function parameters are measured by a fit to the visible part of the shower, usually cut at the beginning (interaction point) and the end (tail absorbed by the earth). All these effects contribute to the systematic errors in
the energy measurement which needs sophisticated monitoring and
calibration techniques, e.g the use of powerful laser beams shot through the
atmosphere. The overall energy resolution one can reach with a
fluorescence telescope is of
course dependent on the EAS energy but also on the detection mode (mono, 
stereo or hybrid). The HiRes detector should have a resolution of 25\% or better above 30~EeV in the \emph{mono} mode.  This improves significantly in the stereo or hybrid modes (about 3\% \emph{median} relative error at the same energy in the latter case).

The identification of the primary cosmic ray with a fluorescence telescope is
based on the shower maximum in the
atmosphere ($X_{\text{max}}$) which depends on the nature and the energy of the incident cosmic ray. At a
given energy, and on the average, a shower generated by a heavy nucleus reaches
its maximum higher in the atmosphere than that of a light nucleus or a proton.
Simulations
show typical values of (respectively for iron nuclei and protons) 750 and 850
g/cm$^2$. Unfortunately, physical fluctuations of the interaction point and
of the shower development (larger than the precision on the shower reconstruction)
blur this ideal image. As an example, at 10~EeV the typical fluctuation on the $X_{\text{max}}$ 
position is 50 g/cm$^2$. Thus, when the fluorescence technique is used alone, it is
practically impossible to define the primary composition on a shower-by-shower
basis.
Therefore, one must look for statistical means of studying the chemical
composition and/or use the hybrid detection method where a multi-variable
analysis becomes possible.

The former method uses the so-called \emph{elongation rate} measured for a
sample of showers within some energy range. The depth of the shower maximum as
a function of the energy for a given composition is given by\cite{Sokolsky}:
$$X_{\text{max}}=D_{\text{el}}\ln\left(\frac{E}{E_0}\right)$$  
where $E_0$ is a parameter depending on the primary nucleus mass. Therefore, 
incident samples of pure composition will be displayed as parallel straight
lines with the same slope $D_{\text{el}}$ (the \emph{elongation rate})
on a semi-logarithmic diagram. The results of such an analysis in the highest
energy range will be mentioned in Section 3.

\subsection{The ground array technique}
\noindent
A ground array is a set of particle detectors distributed as a more or less
regular matrix over some surface. The surface of the array is a
direct function of the expected incident flux and of the statistics needed to
answer the questions at hand. The 100 km$^2$ AGASA array (see Section 3.2) is
appropriate to confirm the existence of the EHECR with energies in excess of 100
EeV (which it detects at a rate of about one event per year). To explore the
properties of these cosmic rays and hopefully answer the open question of their
origin, the relevant detector will no doubt be the Auger Observatory with its
6000 km$^2$ surface over two sites.

The array detectors count the number of secondary particles which cross them
as a function of time. Therefore, they sample the non-absorbed part of the
shower which reaches the ground. The incident CR's direction and energy are
measured by assuming that the shower has an axial symmetry. This assumption is
valid for not too large zenith angles (usually  $\theta<60^{\circ}$). At larger
angles the low energy 
secondaries are deflected by the geomagnetic
fields and the analysis becomes more delicate. 

The direction of the shower axis (hence of the incident CR) is reconstructed by
fitting an analytical function (the ``lateral
distribution function" or LDF) to the measured densities. The LDF explicit form depends on each
experiment. The Haverah Park experiment\cite{Lawrence} (an array of
water-Cerenkov tanks) used the function:
$$\rho(r,\theta,E)=kr^{-[\eta(\theta,E)+r/4000]}$$
as the LDF for distances less than 1 km from the shower core. Here $r$ is in
meters, and $\eta$ can be expressed as:
$$\eta(\theta,E)=a+b\sec\theta+c\log(E/E_0)$$
with appropriate values for all the parameters taken from shower theory and
Monte Carlo studies at a given energy range. At larger distances (and highest
energies), this function has to be modified to take into account a change in
the rate at which the densities decrease with distance. A much more
complicated form is used by the AGASA group.\cite{Yoshida} However, the
principle remains the same.

Once the zenith angle correction is made for the LDF, an estimator of the
primary CR energy is extracted from this function. At energies below 10~EeV,
this estimator is usually taken as the particle density (whatever particles
detected by the array stations) at 600~m from the shower core, $\rho_{600}$.
The density at 600~m is chosen for the following reason.
Because of variations on the primary interaction point (and the position
of the shower maximum), there are large fluctuations in
the ground densities close to the core. At the same time,
the statistical fluctuations in the measured densities are
important at large distances where the densities are low. Monte Carlo
studies show 
that somewhere in between, the overall fluctuation reaches a minimum. This happens
to be at 600~m from the core, a value slowly increasing when one goes to the
highest energies. In the EHECR range, a more appropriate density is
$\rho_{1000}$. Once this value is determined, the primary energy is related to
it by a quasi-linear relation: 
$$E=k\rho_{600}^\alpha$$ 
where $\alpha$ is a parameter close to 1. Of course, to be able to reconstruct
the LDF, many array stations have to be hit at the same time by a shower. The
spacing between the stations determines the threshold energy for a vertical
shower: the 500~m spacing of the Haverah Park triggering stations corresponds
to a threshold of a few $10^{16}$~eV, while the 1.5~km separation of the
Auger Observatory stations makes this array almost 100\% efficient for
energies above 10~EeV.

In a ground array, the primary cosmic ray's identity is reflected in the
proportion of muons among the secondaries at ground level.
Here a proper estimator is therefore the ratio of muons to electrons -~and
eventually photons, if they are detected~- (see Section \ref{ident}). 
When a ground array has muon detecting capabilities (water
Cerenkov tanks, buried muon detectors), one measures directly the muon to
electron ratio. Otherwise, an indirect method consists in measuring the rise-time of
the signal in the detectors: the faster this time, the higher the muon
content.

\subsection{The detection of neutrinos}
\noindent
EHE neutrinos may also be detected by their EAS. This is important for
two reasons. The first is that the detection of  neutrinos (together
with an important component of photons) in the higher energy range of the
spectrum is a solid signature of the top-down mechanisms (see Section
\ref{TopDown}). The second is that the projected high energy neutrino
telescopes (under-water or under-ice km scale detectors) are 
ineffective at energies above $10^{16}$~eV at which the Earth becomes
opaque to upward going neutrinos. Therefore large ground arrays (or fluorescence
telescopes) for which the interaction medium is not the earth but the
atmosphere, and which \emph{could} become efficient enough at $10^{17}$~eV and above, would be complementary to the neutrino telescopes in the
exploration of the whole spectrum. 

The neutrino cross-sections at these ultra-high energies become non-neg\-li\-gible
(about $10^{-32}$~cm$^2$ at 1~EeV).\cite{PAPIERNEUTRINO} The neutrino can
therefore initiate a shower at some point in the atmosphere where the density
is high enough.  The main difficulty for an observer is to identify the EAS as
coming from a neutrino. With the fluorescence technique (which can locate with
some precision the  interaction point), a clear neutrino signature would be 
to see a shower starting deep in the atmosphere.  However, the expected low
fluxes of neutrinos combined with the low cross-sections result in very low
event rates. The 10\% duty cycle of the fluorescence detectors will therefore
hinder their use for this purpose.

An alternative and promising way of detecting ultra-high energy neutrinos would
be to use horizontal air showers,\cite{Billoir} i.e. showers generated by
cosmic rays with incident zenith  angles larger than 60$^\circ$. At these large
angles,  hadronic showers have their electromagnetic part extinguished as they
have gone through a few equivalent vertical atmosphere (2 at 60$^\circ$, 3 at
70$^\circ$, 6 at 80$^\circ$). Only high energy muons survive past 2 equivalent
vertical atmospheres. These muons are  created in the first stages of the
shower development, are of very high energy, and therefore the shape of the
shower front is quite specific: it is very flat (curvature of more than
100~km), and its time extension is very short (less than 50~ns!). On the other
hand, a neutrino shower which would have been initiated a few kilometers before
hitting the array would appear as a ``normal'' shower, with a curved front
(curvature of a few km), a large electromagnetic component, and a signal with a
width of a few microseconds.

With such important differences between neutrino and background (had\-ro\-nic) showers,
particularly at large zenith angles, we can assume that if the fluxes are high
enough, the next generation of experiments will detect neutrino induced
showers. The real background would come, but only at  the highest energies,
from penetrating photon showers whose development is delayed due to  the LPM effect. However, such showers
would show a strong correlation in direction with the geomagnetic field, a
helpful effect to distinguish them from neutrinos. A large amount of ongoing
work is exploring this very promising and exciting field.

\section{What do we know about the EHECR?\label{Auger}} 
\noindent 
\begin{figure}[!ht]
\begin{center}  
\epsfig{file=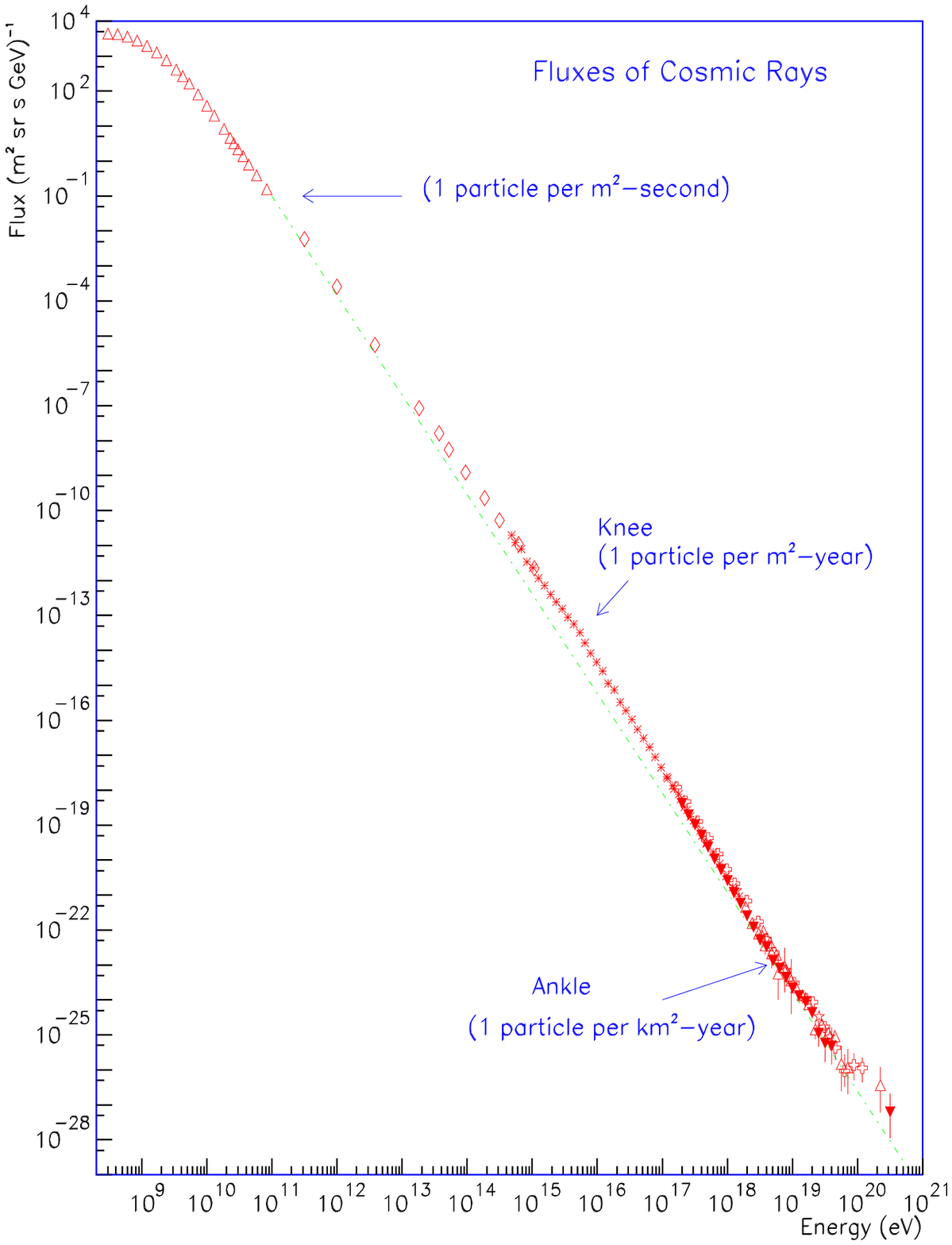,bbllx=86pt,bblly=132pt,bburx=555pt,%
bbury=746pt,width=250pt,clip=}
\end{center}  
\fcaption{The all-particle spectrum of cosmic rays.\cite{Swordy}
\label{spectrum}}  
\end{figure}  
It is outside the scope of this review to present the full history of the
cosmic ray detection and studies. This would cover the whole century (1912 is
the year of the first decisive balloon experiments by Victor Hess). If we want
a starting point for the genesis of the EHECR physics, we need to go back to
the end of the thirties with the first observations of Pierre Auger
 and collaborators.\cite{PAuger} They studied the coincidence
rates between counters with increasing separation (up to 150~m in their first
experiments in Paris, more than 300~m when they repeated them at the
Jungfraujoch in Switzerland). They inferred from this very modest measurement
the existence of primary cosmic rays with energies as large as 1~PeV (\E{15}). Today's
ground arrays used to detect the highest energy cosmic rays (some of them more
than five orders of magnitude beyond the energies Auger considered as being
extraordinary in his times) are actually based on this same technique, however
sophisticated they may be: time coincidences between distant counters to
identify a giant shower, and then the lateral profile of the densities of the
particles hitting the counters to measure the total energy.  

Figure \ref{spectrum} is a compilation\cite{Swordy} of the differential
spectrum of cosmic ray flux as a function of energy. On this figure, integrated
fluxes above three energy values are also indicated: 1 particle/m$^2$-second
above 1~TeV, 1 particle/m$^2$-year above 10~PeV, 1 particle/km$^2$-year above 10
EeV. Ground detectors are the only alternative for the
highest energy part of the spectrum.

In this section, and unless otherwise specified, we shall pay special attention to the 
events with energies exceeding 100~EeV. This value has no particular
physical meaning except that it is well above the GZK cutoff. 

Those events we are interested in (for brevity, let us call them `Big Events')
were observed in very small numbers over the past decades. After a review of
the main experimental setups where their detection took place, we shall present
a short overview of the information one can extract from the available data.

\subsection{Definitions}
\noindent  
The EAS detection performance of a ground based detector is given by its
\emph{acceptance} (or \emph{aperture}) ${\cal A}$. For detectors large enough so
that boundary effects can be neglected, it is assumed that at a given energy an
EAS is detected whenever the shower axis hits the ground within an area $S$
(for most cases the surface covered by the ground array) and with a zenith
angle having a value between $0$ (vertical showers) and  a maximum $\theta_{\text{max}}$, 
usually about $60^{\circ}$. This maximum corresponds either to a loss of acceptance 
(like in ground arrays of thin scintillators) or to a change in the shower properties
as for nearly horizontal showers the interaction point is usually far away from the detector 
and the shower is then mainly composed of fast muons.

The elementary \emph{acceptance}
for a solid angle $d\Omega$ in the direction $\theta$ is defined by: 
$$d{\cal A}=S\cos\theta d\Omega$$
and the total geometrical aperture is:
$${\cal A}=S\int_0^{\theta_{\text{max}}}\cos\theta d\Omega = \pi
S\sin^2\theta_{\text{max}}$$
usually given in km$^2$sr units. One obtains the event rate by convoluting the
 aperture with the incident flux for a given type of particles. 
The geometrical aperture of an optical detector
(like the Fly's Eye) should be multiplied by a reduction factor (of the order of
0.1) due to its duty cycle: such a detector can take data only during clear
moonless nights. In the following, whenever we present the performance of a
fluorescence detector, this factor will be included.

\subsection{Past and present detectors}\label{vr}\label{flyseye}\label{agasa_det}
\noindent  
12 Big Events have been observed in the past\footnote{We do not include here a few 
Big Events detected by the HiRes collaboration (see Section \ref{hires}) 
not yet published.}~ and, after thorough study,
confirmed as having reached or exceeded the energy of 100~EeV. The
corresponding detectors are listed below, with their total exposures in the
relevant energy range.\cite{Watson} 

\bi

  \item {\bf Volcano Ranch}\cite{Linsley} (New Mexico, USA). This is
  the first detector claiming to have detected a Big Event in February 1962. It
  was an array of 3~m$^2$ scintillator counters with a spacing of 900~m and a
  total area of  about 8~km$^2$. The detector's total exposure  was estimated
  to be of the order of 60~km$^2$~sr~year.

  \item {\bf Haverah Park}\cite{Lawrence} (Great-Britain), an array of water
  Cerenkov tanks of various sizes and spacings covering an area of 12~km$^2$.
  The detector took data during almost 20 years (1968-1987) with a total
  exposure of 270 km$^2$~sr~year. It reported 4 Big Events.

  \item {\bf Fly's Eye}\cite{Bird}%\label{flyseye} 
(Utah, USA), a binocular
  system of fluorescence telescopes. It is now being replaced by HiRes, a
  new generation fluorescence detector of larger aperture on the same site. Its
  total exposure is estimated at about 600 km$^2$~sr~year in the mono-mode (an
  EAS detected only by one of the two telescopes) and 170 in the stereo-mode.
  The Fly's Eye detected the most energetic particle ever (320~EeV), but
  unfortunately in the mono-mode, hence with relatively large errors on the
  measurements of the incident direction and energy.

  \item {\bf AGASA}\cite{Yoshida}%\label{agasa_det} 
(Akeno, Japan), the largest,
  and still operating, ground array in the world. It is composed of 111
  scintillator counters (2.2~m$^2$ surface and almost 1~km spacing) over an
  area of 100~km$^2$, with also 27 muon counters. AGASA has taken data since 1990,
  with a total exposure of 670~km$^2$~sr~year. It reported up to now the
  detection of 7 Big Events, including the second in the energy hierarchy at
  200~EeV.

\ei

Another detector (a combination of scintillators, muon and air-Cerenkov
detectors), {\bf Yakutsk}\cite{Afanasiev} (Russia), claims the detection of
a Big Event. However, due to difficulties in estimating the exposure and
energy resolution of this setup at the highest energies, their data are not
included in the flux evaluations.

\subsection{The energy spectrum and flux}
\noindent
The energy spectrum (Figure \ref{spectrum}) ranges over 13 orders of magnitude
in energy and 34~(!) orders of magnitude in flux. However, if one discards the
saturation region at the lowest energies, the spectrum is surprisingly regular
in shape. From the GeV energies to the GZK cutoff, it can be represented
simply by three power-law curves 
%(as is expected if the main accelerating model is the Fermi mechanism\cite{Longair2}) 
interrupted by two breaks, the so-called
``knee" and ``ankle". 
\begin{figure}[!htb]

\begin{center}  
\mbox{ \epsfig{file=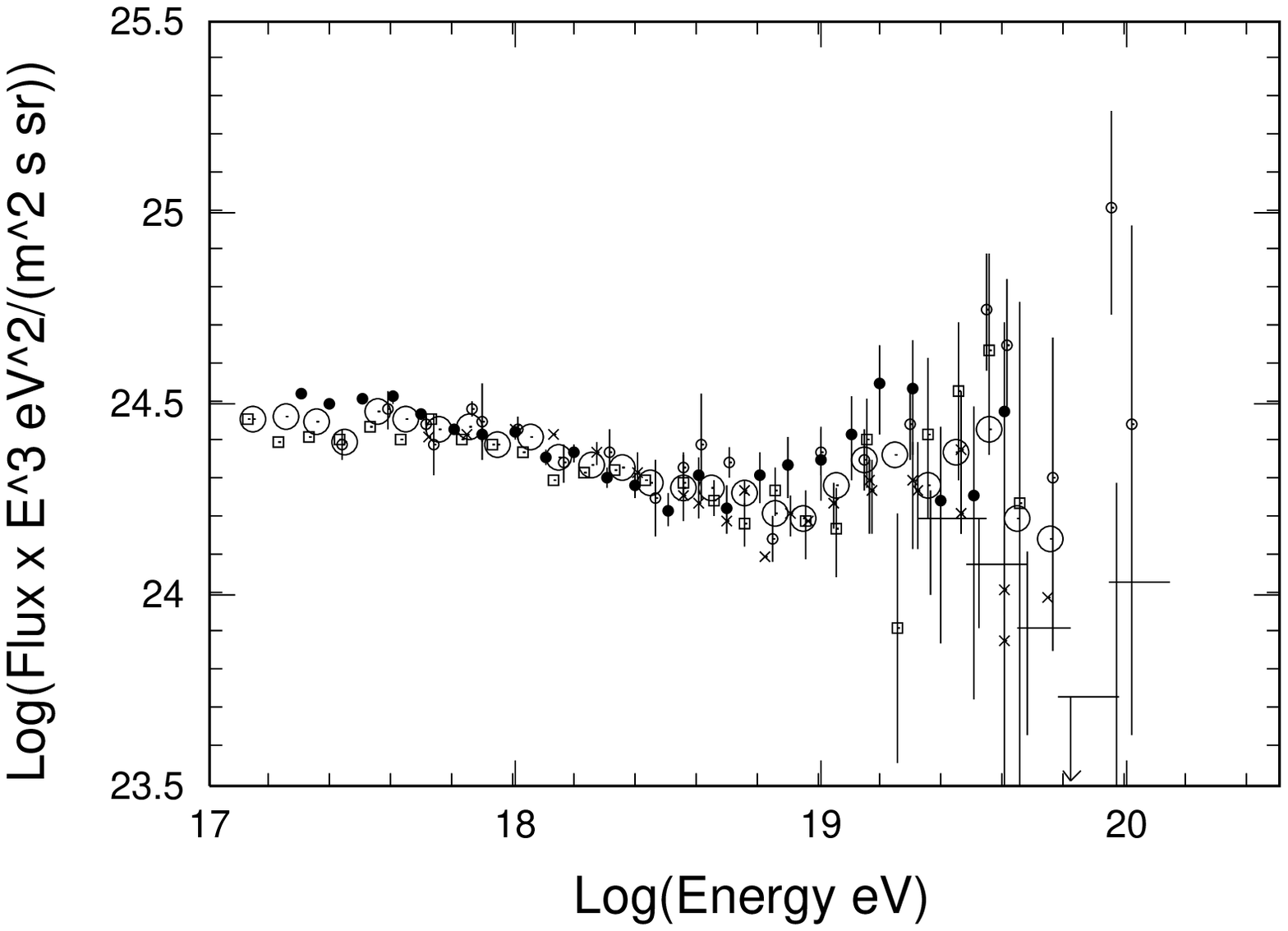,width=270pt}}
\end{center}  
\fcaption{Energy spectrum above 100~PeV. The compilation is from
Ref.\cite{Yoshida} updated by M.Nagano (private communication).\label{agasa}}
\end{figure}

\begin{figure}[!htb]

\begin{center}  
\epsfig{file=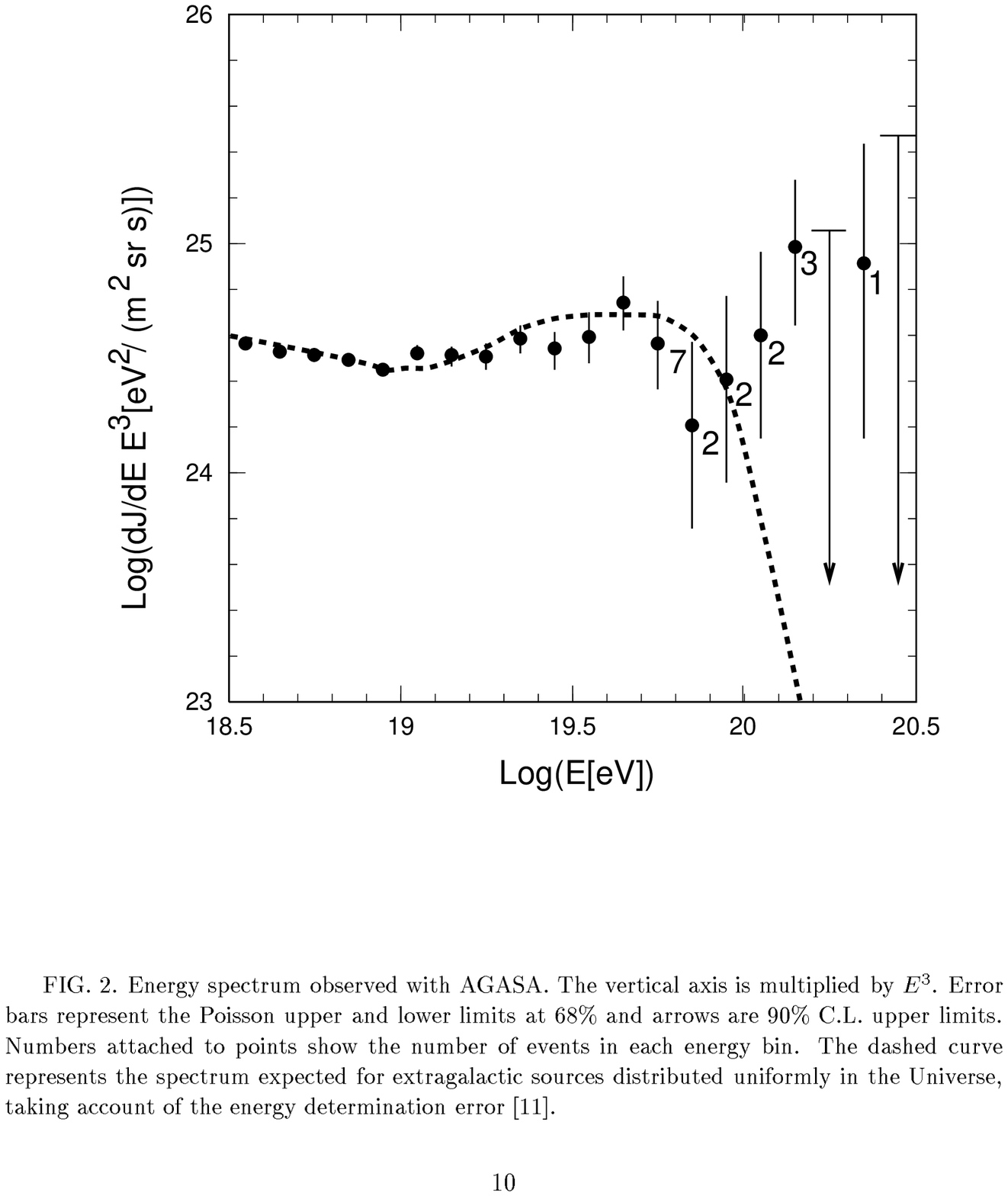,bbllx=117pt,bblly=228pt,bburx=512pt,%
bbury=602pt,width=8cm,clip=}
\end{center}  

\fcaption{Highest energy region of the cosmic ray spectrum as observed by the
AGASA detector.\cite{Takeda} The figures near the data points indicate the
number of events in the corresponding energy bin. The arrows show 90\%
confidence level upper limits. The dashed line is the expected spectrum if the
sources were cosmologically distributed.\label{takeda}}
\end{figure}

The supra-GZK events, and especially those above 100~EeV mentioned in the
preceding section, were submitted to very close scrutiny, especially in the
recent years where several dedicated projects were proposed to study this
energy region. Their flux is extremely low. Figure \ref{agasa} is a zoom on the
highest energy range of the spectrum. On this figure, 
the energy spectrum is multiplied by $E^3$ so
that the part below the EeV energies becomes flat. One can see the `ankle'
structure in its complexity: a steepening around the EeV and then a confused
region where the GZK cutoff is expected. The ultimate data points come from
very few events hence their large error bars, and due to normalization problems it is
difficult to compare different experiments. On Figure \ref{takeda} where the
AGASA data alone are displayed,\cite{Takeda} one has a clearer view of what is
expected (within a conventional framework) and what is observed. 
Here, the GZK cutoff is shown by the dashed line which is the expected spectrum
shape with the hypothesis of a cosmologically uniform distribution of the
sources. The data suggests a change of slope as if 
a new phenomenon was rising above a steeply falling spectrum.

The flux of the highest energy cosmic rays obviously cannot be deduced from the
data above the cutoff energies: no reliable fit to the spectrum shape is
feasible in this region. A reasonable way of defining an order-of-magnitude
flux is to base one's estimation on the total exposures mentioned above and the
number of events observed. The exposure to events above the GZK cutoff for AGASA, Fly's Eye and Haverah
Park detectors together, is of the order of 2000
km$^2$~sr~yr. The number of events observed in this energy range over
the lifetime of these detectors would yield an integrated flux which can be
roughly parametrized by\label{flux}:

\begin{equation}
I(E>E_0)\approx\left(\frac{E_0}{10\,\text{EeV}}\right)^{-2}\,
\text{km}^{-2}\text{sr}^{-1}\text{year}^{-1}\label{eq:flux}
\end{equation} 
 With $E_0=100$
EeV, the expected flux is 1 particle per km$^2$ per century! A
statistically significant sample of events (necessary to reconstruct the shape
of the spectrum, locate eventual point sources, study possible
anisotropies in the incoming directions, and have hints on the chemical
composition of the EHECR) cannot be obtained within a reasonable time unless
the detectors have huge apertures. A typical and necessary order of magnitude
would be $10^4$~km$^2$~sr~yr, ensuring a hundred events per year above 100
EeV and ten thousand above 10~EeV. 

The GZK cutoff, as it would be expected if the sources are cosmologically
distributed and if the observed cosmic rays have no exotic propagation and interaction properties, \emph{is} violated. This statement, based on experimental data, can hardly be questioned. The next generation of detectors we will describe in Section 5 are designed to help us understand why and how.

\subsection{The chemical composition}
\label{Compo}
\noindent
The methods to identify the primary cosmic
ray from the EAS parameters were described in Section 2. 
One must keep in mind that observables contributing to the
identification (depth of the shower maximum, muon to electron-photon ratio in
the EAS at a given depth, position of the first interaction...) are subject to
large fluctuations and each identity signature includes an unavoidable
background. Therefore, the EHECR chemical composition is very likely to be
unveiled only on a statistical basis, and by cross-checking the information
deduced from as large a number of parameters as possible. 

What we know at present on this issue is weak and controversial. The most
recent information comes from the Fly's Eye and AGASA experiments. The  Fly's
Eye studies\cite{Bird} (between 0.1~EeV and 10~EeV) are based on $X_{\text{max}}$ 
behavior as a function of the logarithm of the primary energy.  With this
method, their data show evidence of a shift from a dominantly heavy composition
(compatible with iron nuclei) to a light composition (protons), so that, if
this interpretation is to be believed, the EHECR would mainly be protons.

The AGASA group based their primary identification on the
muon content of the EAS at ground level,\cite{Hayashida} an essentially
independent method. Showers initiated by heavier nuclei are expected to have a
higher muon content than light nuclei or protons, the
gamma showers being the poorest in muons. Initially the conclusion of the AGASA
experiment was quite opposite to the Fly's Eye: no
change in chemical composition. However a recent critical review of both
methods\cite{Dawson} showed that the inconsistencies were mainly due to the
scaling assumption of the interaction model used by the AGASA group. 
The authors concluded that if a model with a higher (compared to the one given by scaling) 
rate of energy dissipation at high energy is assumed, as indicated by the direct $X_{\text{max}}$ 
measurements of the Fly's Eye, both data sets demonstrate a change of composition, a shift 
from heavy (iron) at 0.1~EeV to light (proton) at 10~EeV. Different interaction models as long
as they go beyond scaling, would lead to the same qualitative result but eventually with a 
different rate of change.

Gamma rays have also high cross sections with air and are still another possible candidate 
for  EHECR but no evidence were found up to now for a gamma signature among the Big Events. 
The most energetic Fly's Eye event was studied in detail\cite{Halzen2} and found incompatible with
an electromagnetic shower. Both interpretations of the AGASA and the Fly's Eye data 
favor a hadronic origin.

\subsection{Distribution of the sources} 
\noindent 
A necessary ingredient in the search for the origin of the EHECR is to locate
their sources. This is done by reconstructing the incident cosmic ray's
direction and checking if the data show images of point sources or correlations
with distributions of astrophysical objects in our vicinity. Since a likely
possibility as for the nature of the EHECR is that they are protons, we will
say a few words about what we know of the galactic and extragalactic magnetic
fields: this will show that for supra-GZK energies, proton astronomy is
possible to some extent. We will then give a review of what we can extract from
the present data in terms of anisotropy of the reconstructed directions and
observed multiplets of events one may consider as images of point sources.

\subsubsection{Magnetic fields}\label{MagField} 
\noindent 
There are a limited number of methods to study the magnetic fields on 
galactic or extragalactic scales.\cite{Kronberg} One is the measure of the
Zeeman splitting of radio or maser lines in the interstellar gas. This method
informs us mainly on the galactic magnetic fields, as extragalactic signals suffer Doppler smearing
while the  field values are at least three orders of
magnitude below the galactic ones. The 
magnetic field structure of the galactic disc is therefore thought to be rather well
understood. One of the parametrizations currently used is that of Vall\'ee\cite{Vallee}: concentric
field lines with a few $\mu$G strength and a field reversal at about one half of
the disk radius. Outside the disk and in the halo, the field model is based on
theoretical prejudice and represented by rapidly decreasing functions (e.g. gaussian
outside the disk).

The study of extragalactic fields is mainly based on the Faraday rotation
measure (FRM) of the linearly polarized radio sources. The rotation angle
actually is a measurement of the integral of the function $n_eB_{\parallel}$
where $n_e$ is the electron/positron density and $B_{\parallel}$ the
longitudinal field component along the line of sight. Therefore, the FRM
needs to be complemented by another measurement, namely that of $n_e$. 
This is done
by observing the relative time delay versus frequency of waves emitted by a
pulsar. Since the group velocity of the signal depends simultaneously on its
frequency but also on the plasma frequency of the propagation medium,
measurement of the dispersion of the observed signals gives an upper limit on
the average density of electrons in the line of sight. 
Here again, because of the faintness of extragalactic signals, our knowledge of
the strength and coherence distances of large scale extragalactic fields is
quite weak and only upper limits over large distances can be extracted. An
educated guess gives an upper limit of 1 nG for the field
strength and coherence lengths of the order of 1 Mpc.\cite{Kronberg} 

\begin{figure}[htbp]

\begin{center}  
\epsfig{file=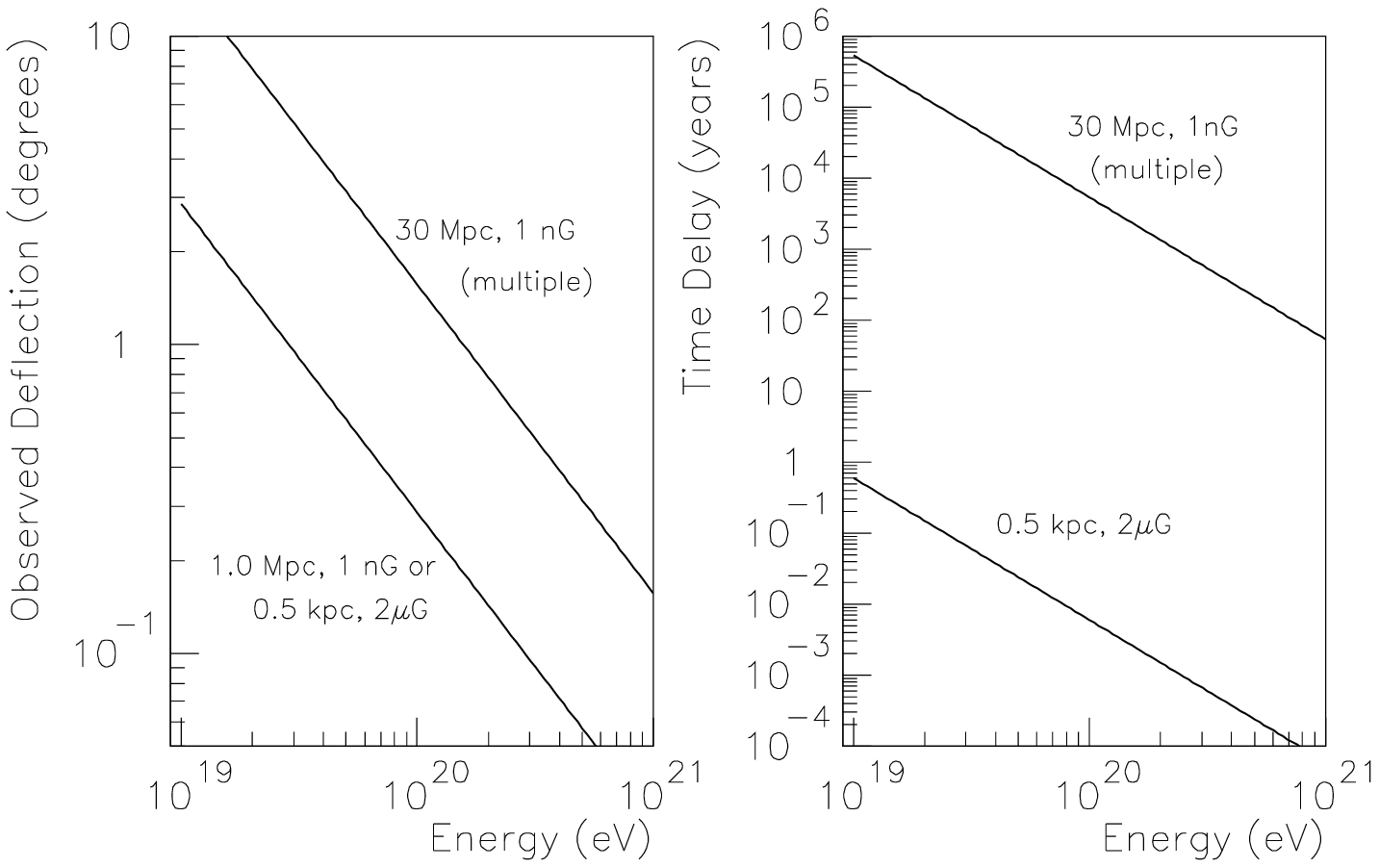,height=214pt,width=322pt}
\end{center}  

\fcaption{Effect of magnetic fields on the propagation of a proton as a
function of energy: angular deviation (left) and time delay (right), with
respect to a straight line trajectory, in the framework of three realistic
scenarios (see text).\cite{Auger}\label{angular}}
\end{figure}
A few other more or less indirect methods exist for the study of large scale
magnetic fields. \emph{If} the EHECR are protons and \emph{if} they come from
point-like sources, the shape of the source image as a function of the cosmic
ray energy will certainly be one of the most powerful of them. The Larmor radius
$R$ of a charged particle of charge $Ze$ in kiloparsecs is given by:
$$R_{\text{kpc}}\approx\frac{1}{Z}\left(\frac{E}{1\,\text{EeV}}\right)
\left(\frac{B}{1\,\mu\text{G}}\right)^{-1}$$
The Larmor radius of a charged particle at 320~EeV is larger than the size of
the galaxy if its charge is less than 8. If we take the currently accepted
upper limit ($10^{-9}$~G) for the extragalactic magnetic fields, a proton of
the same energy should have a Larmor radius of 300~Mpc or more.

In Figure \ref{angular}, three different situations are envisaged to evaluate
the effects of magnetic fields on a high energy cosmic proton. The situations
correspond to what is expected a-) for a trajectory through our galactic disk
(0.5~kpc distance inside a 2~$\mu$G field) or b-) over a short distance (1~Mpc)
through the extragalactic (1~nG) field (same curve), and finally c-) a 30~Mpc
trajectory through extragalactic fields with a 1~Mpc coherence length (multiple
scattering effect). One can see that at 100~EeV, the deviation in the third
case would be about $2^{\circ}$. This gives an idea of the image size if the
source is situated inside our local cluster or super-cluster of galaxies.
Since the angular resolution of the (present and future) cosmic ray detectors
can be comparable to or much better than this value, we expect to be able to
locate the point-like sources (if they exist) or establish correlations with
large-scale structures. 

However, let us remember that this working hypothesis of very weak extragalactic
magnetic fields is not universally accepted. Several authors recently advocated
our bad knowledge of those fields showing that the arguments which will be
developed in the following sections (the puzzling absence of correlations
between the direction of the EHECR and either point sources or large
structures) drop out if one envisages stronger magnetic fields (typically at the
$\mu$G level) either locally\cite{Lemoine} or distributed over larger,
cosmological, scales.\cite{Farrar2}\label{Xfield}

\subsubsection{Anisotropies}
\noindent
\begin{figure}[!htb]

\begin{center}  
\epsfig{file=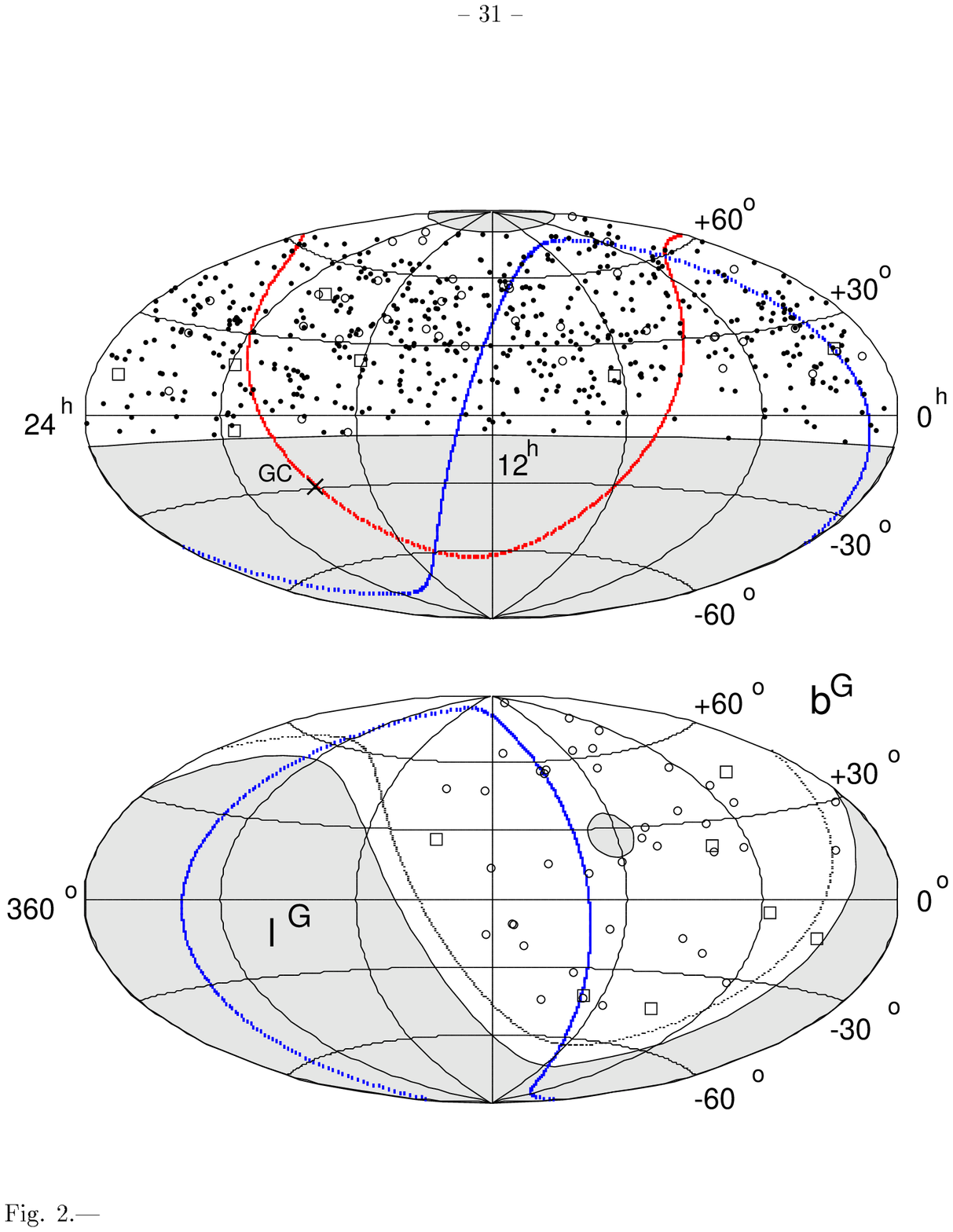,bbllx=79pt,bblly=417pt,bburx=528pt,%
bbury=627pt,width=12cm,clip=}
\end{center}  

\fcaption{Arrival directions of cosmic rays with energies above 10~EeV
(equatorial coordinates), as measured by the AGASA experiment.\cite{Takeda2}
The thick dotted lines show the galactic and supergalactic planes (GC
indicating the galactic center). The shaded regions are those invisible to the
AGASA detector. See text.\label{bigevents1}}
\end{figure}
In the search for potential sources, the propagation arguments incite us to
look for correlations with the distribution of astrophysical matter within a
few tens of Mpc. In our neighborhood, there are two structures
showing an accumulation of objects, both only partially visible from any
hemisphere: the galactic disk on a small scale and the supergalactic plane on a large scale, a
structure roughly normal to the galactic plane, extending to distances up to
$z\approx 0.02$ (about 100 Mpc) and correlated with a denser distribution of
radio galaxies. In equatorial
coordinates with the hypothesis of an isotropic distribution of the sources, and
over large periods of data taking, the right ascension distribution of events
must be uniform and the declination distribution can be parametrized with the
known zenith angle dependence of the detector aperture. 

The most recent analysis on the correlation between arrival directions and
possible source locations was done by the AGASA experiment for the highest energy range.\cite{Takeda2} The
analysis is based on 581 events above 10~EeV, a subset of 47 events above 40
EeV and 7 above 100~EeV. Figure \ref{bigevents1} is a compilation of the total sample in equatorial
coordinates. The dots, circles and squares are respectively events with
energies above 10, 40 and 100~EeV. The data show no
deviation from the expected uniform right ascension distribution. An excess of
2.5~$\sigma$ is found at a declination of $35^{\circ}$ and can be interpreted as
a result of observed clusters of events (see below). No convincing deviation
from isotropy is found when the analysis is performed in galactic coordinates.

The same collaboration\cite{Hayashida2} made also a similar analysis for the lower energy region (events
down to 1~EeV and detected with zenith angles up to $60^{\circ}$). In this
article, a slight effect of excess events in the direction of the galactic
center was announced. A similar study\cite{Bird2} with the Fly's Eye data, concludes on a small
correlation with the galactic plane for events with energies lower than 3.2
EeV and isotropy for higher energies.

In summary, both the AGASA and Fly's Eye experiments seem to converge on some anisotropy in the EeV range (correlation with the galactic plane and center) and isotropy above a few tens of EeV. This result may seem surprising - one
naively expects the correlations to be stronger when the cosmic rays have large
magnetic rigidity. It is actually explained by the fact that the low energy
component may be dominantly galactic heavy nuclei (see the section on chemical
composition), hence a (weak) correlation with the galactic plane, whereas the
higher energy cosmic rays would be dominated by extragalactic protons. 

With the present data it is difficult to come to any clear conclusion on the ani\-so\-tro\-py issues, especially at the highest energies. 
All we can do is acknowledge the problems and make a list of
technical requirements to solve them. First of all, and whatever the analysis
methods used, statistics is the sinews of war. One needs at least an order of
magnitude in the rate with which we are detecting the relevant events in the
highest energy range, hence in the collecting detectors' aperture. This is the
principal aim of the next generation of experiments. Second, the powerful
harmonic analysis needs full sky coverage and, as much as possible, uniform
exposure. This means data from both hemispheres. All the data used in the above
analyses come from the northern hemisphere. Possible correlations with the
galactic plane or the galactic center will only be confirmed, or invalidated,
if a detector systematically explores the southern sky. Also,
due to day-night effects, a fluorescence detector does not have a uniform
right-ascension coverage. One should also take into account the angular
acceptance. The scintillator technique (used in the AGASA detector) has an
intrinsic limitation for large angles. The use of Cerenkov tanks (as in the
Haverah Park experiment) circumvents this limitation: it can even detect
horizontal showers with equal efficiency, although the bending of the charged
particles of the shower makes the analysis more difficult for large angles. All
these constraints have been carefully considered in the design of the next generation
experiments.

\subsubsection{Point sources?} 
\noindent If the sources of EHECR are nearby astrophysical objects and if, as
expected, they are in small numbers,
a selection of the events with the largest magnetic rigidity would combine
into multiplets or clusters which would indicate the
direction to look for an optical or radio counterpart. Such
an analysis was done systematically by the AGASA group.\cite{Takeda2}  
\begin{figure}[!htb] 

\begin{center}  
\epsfig{file=bigevents.eps,bbllx=71pt,bblly=187pt,bburx=528pt,%
bbury=398pt,width=12cm,clip=} 
\end{center}   
 
\fcaption{Arrival directions of cosmic rays with energies above 40~EeV
(galactic coordinates), as measured by the AGASA experiment.\cite{Takeda2} See
text.\label{bigevents2}} 
\end{figure} 
Figure \ref{bigevents2} shows the subsample of events in the AGASA catalog with
energies in excess of 100~EeV (squares) and in the range 40-100~EeV (circles).
A multiplet is defined as a group of events whose error boxes ($2.5^{\circ}$
circles) overlap. One can see that there are three doublets and one triplet. If
one adds the Haverah Park events, the most southern doublet also becomes a
triplet. The chance probability of having as many multiplets as observed with
a uniform distribution are estimated by the authors to less than 1\%.\footnote{The chance probability is very difficult to evaluate in an \emph{a posteriori} analysis and depends 
strongly on the assumed experimental error box size.}

A search for nearby astrophysical objects within an angle of $4^{\circ}$ from any event
in a multiplet was also done, and produced a few objects. One of the most
interesting candidates is Mrk 40, a galaxy collision, since the shock
waves generated in such phenomena are considered by some authors\cite{Cesarsky}
as being valid accelerating sites.

Another way of using the observed multiplets, \emph{assuming} that they come
from an extragalactic point source, is to consider the galactic disk as a 
magnetic spectrometer which can give information on the charge of the incident
cosmic rays. Cronin\cite{Cronin} made such an analysis on the doublet where
the energy difference between the two events is the largest (a factor of four).
He uses the magnetic field model of Vall\'ee\cite{Vallee} to trace back the
detected couple of events outside of the galaxy assuming various charges. It
is shown that the maximum charge compatible with a separation less than the
detector's angular resolution is 2 for the members of the doublet with
conservative integrated values for the magnetic field, a result 
compatible with protons being the most likely EHECR.

\subsubsection{Is a galactic origin possible?}
\noindent
One can put forward at least three reasons to argue for an extragalactic origin of the EHECR sources:
\begin{enumerate}
\item  The galactic
size and magnetic fields are such that the EHECR cannot be confined inside the galaxy;
\item One usually assumes that there are no likely
astrophysical objects which can be considered as remarkable accelerators inside
our galaxy; 
\item \emph{If} the accelerated particles are light
nuclei, their direction should show correlations with the galactic plane, which
is not the case. 
\end{enumerate}

The second argument was sometimes challenged in the past by authors who propose some galactic objects
such as microquasars and pulsars as possible powerful accelerating engines. None 
of these ideas seem to us fully convincing, in the sense that they are generally based on \emph{ad hoc} 
hypotheses and fail to explain the full set of observational data on the EHECR (see 
e.g. Blanford\cite{Blanford} for a recent review). 

\section{Possible sources of the EHECR}
\noindent
Today's understanding of the phenomena responsible for the production of EHERC, i.e. the transfer of macroscopic amounts of energy to 
microscopic particles, is still limited.
One distinguishes two classes of processes: the so called ``Top-Down''
and ``Bottom-Up'' scenarios. In the former,
the cosmic ray is one of the stable decay products of a 
super-massive particle. Such particles 
with masses exceeding 1~ZeV can either be meta-stable relics of some primordial field or highly 
unstable particles produced by the radiation, interaction or collapse of topological defects. 
Those processes are reviewed in Section~\ref{TopDown}

In the second scenario discussed in section~\ref{BottomUp} the energy is transferred 
to the cosmic rays through their interaction with electromagnetic
fields. This classical approach does not require new physics as opposed to the ``Top-Down'' mechanism, but does not exclude it either since, in some models, the accelerated particle - the cosmic ray -
is itself ``exotic''. 

Once accelerated the cosmic rays must propagate from their source to the observer. 
At energies above 10~EeV and except for neutrinos, the Universe 
is not transparent to ordinary stable particles on scales much larger than about 10 Mpc. 
Regardless of their nature, cosmic rays lose energy in their interaction with the various photon backgrounds, 
dominantly the copious Cosmic Microwave Background (CMB) but also the Infra-Red/Optical (IR/O) and 
the Radio backgrounds. 
The GZK cutoff puts severe 
constraints on the distance that a cosmic ray can travel before losing most of its energy or being 
absorbed. The absence of prominent visible astrophysical objects in the direction of the observed highest 
energy cosmic rays together with this distance cutoff adds even more constraints on the ``classical'' 
Bottom-Up picture.

It is beyond the scope of this review to describe all the scenarios - they are
far too numerous - proposed for the production of the EHECR. Let us simply agree on the fact that the profusion of models shows that none of them is totally satisfactory. 
Consequently we will try to present, from an experimentalist's point of view, 
the main features of the various categories of models.
We will also try to focus on the possible experimental constraints, if they exist, or on the 
problems related to the EHECR and which remain unanswered. For a more detailed review we urge the reader to
consult the excellent report by P.Bhattacharjee and G.Sigl\cite{Batt_Sigl} and the references therein. Extensive use of this report is made in some of the following sections where we avoided repeated reference to it. 

At first sight, it would seem natural to discuss potential sources and acceleration mechanisms before
the description of the cosmic ray transportation to Earth. However, and because the attenuation or 
interaction lengths are relatively
short and strongly energy dependent in the range of interest, the observed spectra do not only depend on the
nature of the sources  but also on their distribution. In addition, the GZK cutoff 
puts important constraints which we prefer to discuss before describing the possible 
nature of the sources themselves.

\subsection{Propagation}
\noindent
We will focus here on the propagation of atomic nuclei (in particular protons) and photons. 
Electrons are not considered as potential EHECR because they radiate most of their energy while 
crossing the cosmic magnetic fields.  Among the known stable particles, and within the framework of the Standard Model, those are
the only possible candidates for EHECR. As we mentioned in Section~\ref{Compo}, the 
actual data effectively favor a hadronic composition. 

Neutrinos and the lightest super-symmetric particles (LSP) should deserve special attention as they may travel
through space unaffected even on large distances.
However, for neutrinos the interaction should occur uniformly in atmospheric depth,   
a feature which is not reproduced by the current data. 
While neutrinos may very well be one of the components of the high energy end of the cosmic ray 
spectrum and prove to be an unambiguous signature of the new physics underlying the 
production mechanisms (see below) they do not seem to dominate the observations at least up 
to energies of a few \E{20}. 

The LSP are expected to have smaller interaction cross sections with photons and a higher threshold for pion photo-production due to their higher 
mass (see Eq.~(\ref{eq:GZK}) below). Therefore they may travel unaffected by the CMB on distances 
10 to 30 times larger than nucleons. However in usual models the LSP is neutral and cannot be 
accelerated in a Bottom-Up scenario and must be produced as a secondary of an accelerated charged 
particle (e.g. protons). This accelerated particle must reach energies 
at least one order of magnitude larger than the detected energy (order of ZeV) and will produce photons. 
The acceleration site should therefore be detectable as a 
very powerful gamma source in the GeV range. In a Top-Down scenario including Super-Symmetry, 
the problem of propagation is of somewhat lesser importance as the decaying super-massive particles 
may be distributed on cosmological or on nearby scales and are, in any case, 
invisible (see Section~\ref{TopDown}). Finally let us stress that the analysis of the 
EHECR shower shape limits the mass 
of the cosmic ray to about 50~GeV,\cite{Albuquerque} an additional  constraint for the LSP candidate.

\subsubsection{Protons and Nuclei: The GZK cutoff}\label{Toto}\noindent 
The  Greisen-Zatsepin-Kuzmin (GZK) cutoff (see Section~\ref{intro}) threshold for  
collisions between the cosmic microwave background (CMB) and protons (photo-pion production) 
can be expressed in the CMB ``rest'' frame as 

%The photon energy threshold for photo-pion production expressed in the proton rest 
%frame is 
%$$ E_\gamma(E_p = m_p) = \frac{m_\pi^2}{2m_p} + m_\pi  \simeq 150\, \text{MeV}$$   
%which translates in the ``laboratory'' frame as
\begin{equation}\label{eq:GZK}
 E_{th} \simeq \frac{E_\gamma m_p}{2\epsilon} \sim \frac{7\times 10^{16}}{\epsilon}
\end{equation}
where $\epsilon$ is the target photon energy and $E_{th}$ the proton threshold, both in electron-volts. 
For an energetic CMB photon with $\epsilon=$ \E{-3}, $E_{th}$ is $7 \times $\E{19} which is where one expects the GZK 
cutoff to start.
 
The interaction length for this process can be estimated from the photo-pion cross section (taken beyond the $\Delta$
resonance production) and the CMB photon density:
$$ L = (\sigma \rho)^{-1} \simeq 1.7\times10^{25}\,\text{cm} \simeq 6\, \text{Mpc} $$        
for $\rho = 410\,\text{cm}^{-3}$ and $\sigma = 135~\mu$barns. 

\begin{figure}[!htb]

\begin{center}  
\epsfig{file=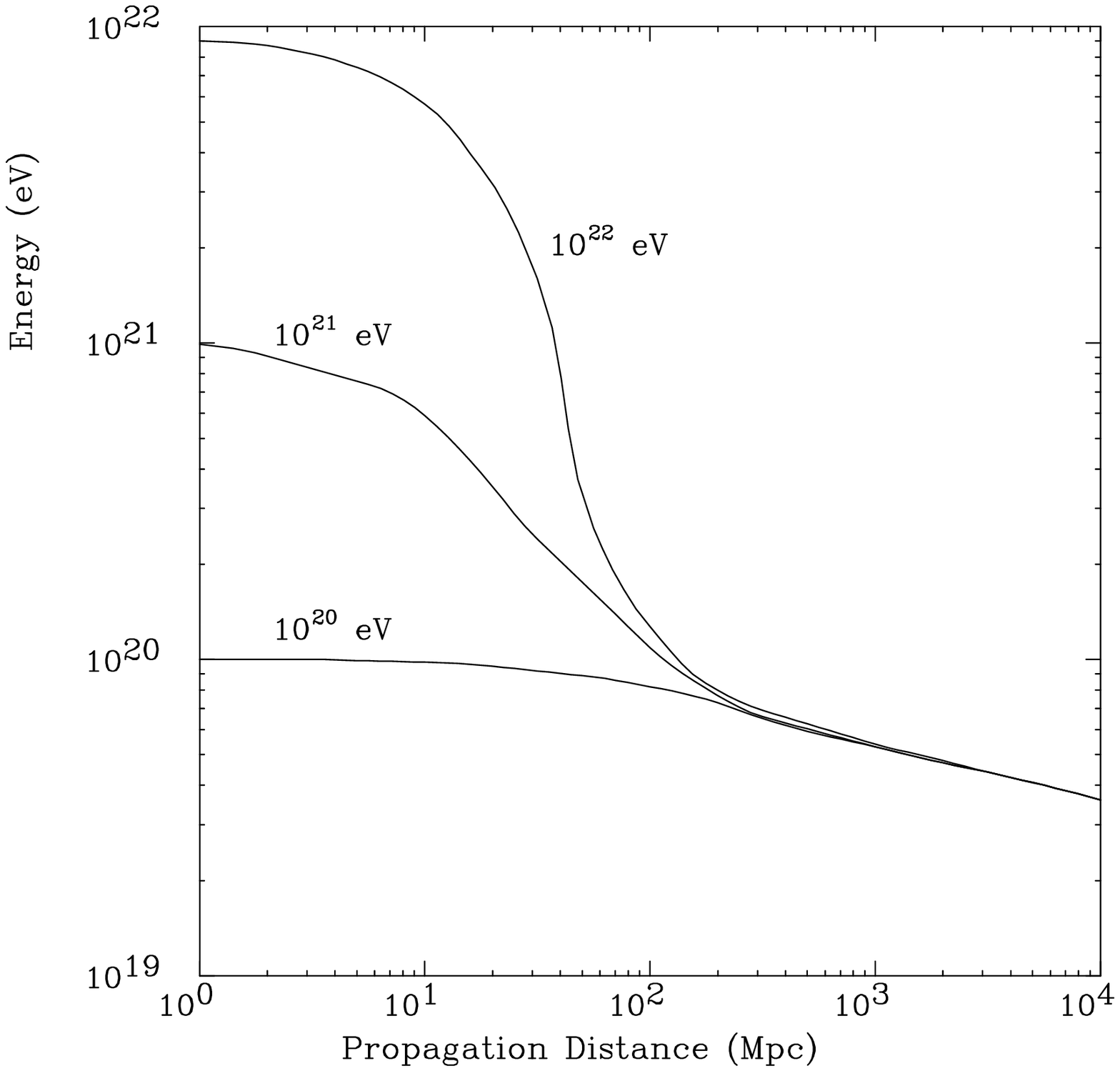, width = 8cm}
\end{center}

\fcaption{Energy of a proton as a function of the propagation distance through the 2.7K cosmic background  
radiation for various initial energies.\cite{Auger}\label{adrf25}}
\end{figure}

The energy loss of protons of various initial energies as a function of the propagation 
distance is shown in Figure~\ref{adrf25}. Above 100 Mpc the observed energy is below 
\E{20} regardless of its initial value. One should point 
out that this reduction is not the consequence of a single catastrophic process but of many 
collisions (more than 10) 
each of which reduces the incident energy by 10 to 20\%.
Therefore the probability to travel without losses is negligible.

A proton may also produce $e^+e^-$ pairs on the CMB at a much lower threshold (around $5\times$\E{17}) but the 
cross section is orders of magnitude smaller and together with a smaller inelasticity 
the overall interaction length stays around 1~Gpc.

\begin{figure}[!htb]

\begin{center}  
\epsfig{file=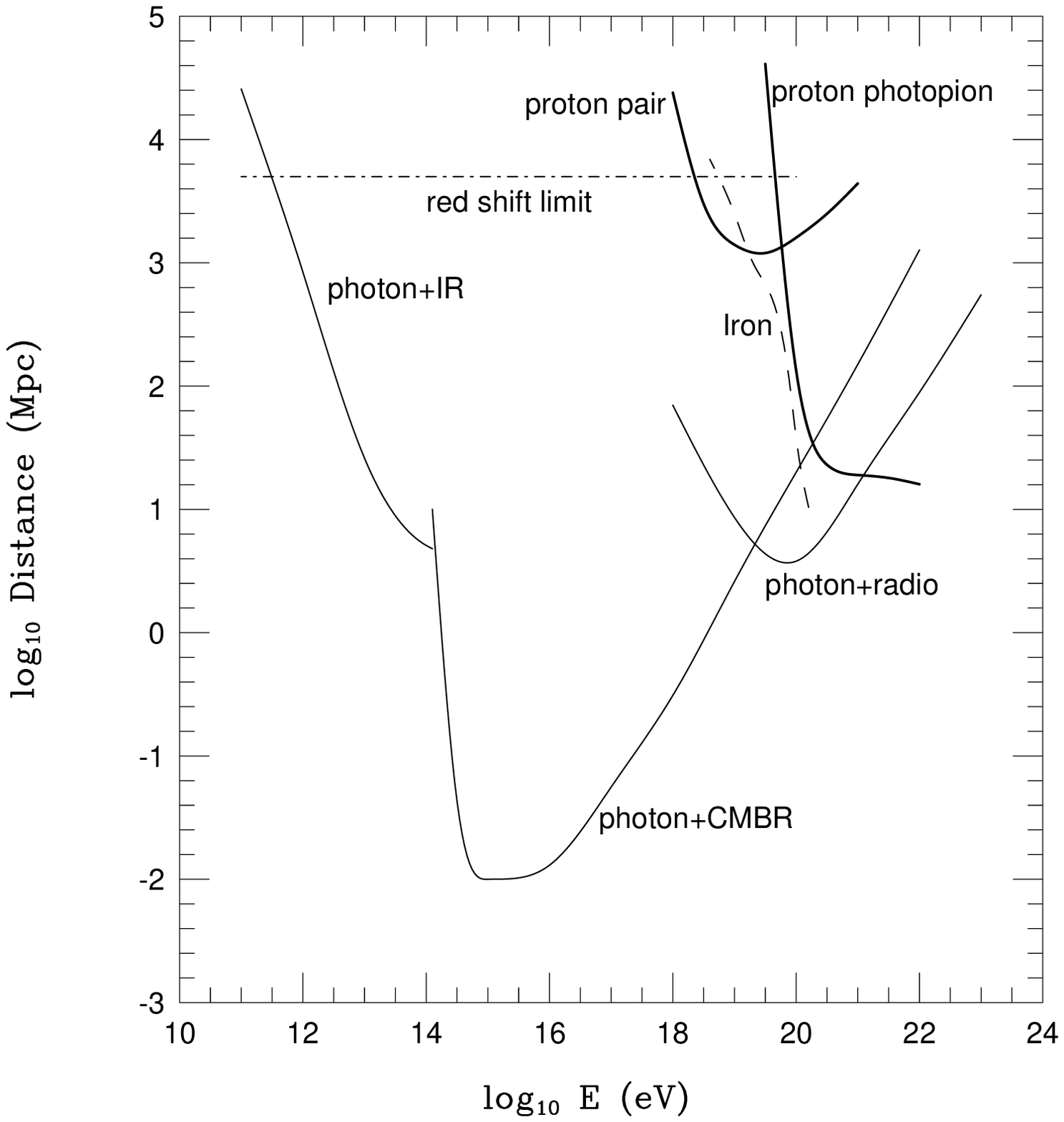, width = 8cm}
\end{center}

\fcaption{Attenuation length of photons, protons and iron in various background radiations as a function of 
energy.\cite{Auger} The dot-dashed line represents the absolute upper limit on the distance a 
particle can travel toward Earth, regardless of its initial energy.\label{adrf24}}
\end{figure}

For nuclei the situation is in general more difficult. 
They undergo photo-dis\-in\-te\-gra\-tion in the CMB and infrared radiations
losing on average 3 to 4 nucleons per Mpc when their energy exceeds 2$\times$\E{19} to 2$\times$\E{20} depending on the IR background density value. 
The IR background is much less well known than the CMB and the attenuation length 
(see Figure~\ref{adrf24}) derived for nuclei must be taken with precaution. 
  
\subsubsection{Electrons and Photons: Electromagnetic Cascades\label{photon}}\noindent
Top-Down production mechanisms predict that, at the source, photons (and neutrinos) 
dominate over ordinary hadrons
by about a factor of ten. An observed dominance of gammas in the supra-GZK range would then be an almost 
inescapable signature of a super-heavy particle decay. Photons are also secondaries of more 
ordinary processes such as photo-pion production; their propagation is thus worth studying.
Unlike photons, electrons and positrons cannot constitute the  primary CR as the radiation energy losses they 
undergo forbid them to reach the highest energies by many orders of magnitude. 

High energy photons traveling through the Universe produce $e^+e^-$ pairs when colliding with the 
Infra-Red/Optical (IR/O), CMB or Universal Radio Background (URB) photons. As can be seen on 
Figure~\ref{adrf24} the attenuation length gets below 100 Mpc for photon energies between $3\times$\E{12} 
and \E{22}. In this energy range, nearly 10 orders of magnitude, the Universe is opaque
to photons on cosmological scales.

Once converted, the $e^+e^-$ pair will in turn produce photons mostly via Inverse Compton Scattering (ICS) 
(the case of synchrotron radiation, usually non dominant, will be treated in the next section). At our energies,
those two dominant processes  are responsible for the production of electromagnetic (EM) cascades.
  
Much above the pair production threshold ($s \gg 4 m_e^2$, where $\sqrt{s}$ is the CM energy) 
the ICS ($\sigma_{\text{ICS}}$) and the pair production ($\sigma_{\text{pp}}$) 
cross sections are related by~:
$$ \sigma_{\text{pp}} \approx 2 \sigma_{\text{ICS}} = \frac{3}{2} \sigma_T \frac{m_e^2}{s} \log\left ( \frac{s}{2m_e^2}\right )$$
where $m_e$ is the electron mass and 
$\sigma_T=8\pi\alpha^2/3m_e^2=665$~mbarn the Thomson cross section of photon
elastic scattering on an electron at rest. The $1/s$ dependence implies that far from the pair  
threshold the EM cascade develops slowly as it is the case when the initial photon energy is above \E{22}. 

At the pair production threshold ($s \sim 4 m_e^2$), the pair cross section rea\-ches a value of \mbox{$\sim170$~mbarn} and 
$\sigma_{\text{ICS}}$  is nearly equal to the Thomson cross section. The EM cascades develop very rapidly. From 
Figure ~\ref{adrf24} one sees that at the pair production threshold against the CMB photons 
($2\times$\E{14}) conversion occurs on distances of about 10~kpc (a thousand times smaller than for protons at 
GZK energies) while subsequent ICS of electrons on the 
CMB in the Thomson regime will occur on even smaller scales (1~kpc). 

As a consequence, all photons of high energy (but below \E{22}) will produce, through successive collisions
on the various photon backgrounds (URB, CMB, IR/O), lower and lower energy cascades 
and pile up in the form of a diffuse 
photon background below \E{12} with a typical $\alpha=1.5$ power law spectrum.  This is a very important fact
as measurements of the diffuse gamma ray background in the $10^7$-\E{11} range done for example by EGRET\cite{EGRET} 
will impose limits on the photon production fluxes of Top-Down mechanisms and consequently on the abundance of 
topological defects or relic super-heavy particles.

\subsubsection{Charged Particles: Magnetic Fields}\noindent
The effect of magnetic fields (galactic or extragalactic) on the deflection of charged particles has been 
reviewed in Section~\ref{MagField}. Here we will present some of the effects of the fields on EM cascade 
production.

Electrons and positrons produced through EM cascades lose energy via synchrotron radiation at a rate given by:

$$ -\frac{dE}{dt} = \frac{4\alpha^2}{3 m_e^2}B^2\left(\frac{E}{m_e}\right)^2,$$ 
where we assume a random field $B$ isotropically distributed with respect to the 
electron direction.

At high enough energy, i.e.
$$ E \sim \left(\frac{B}{10^{-9}\,\text{G}} \right)^{-1}10^{19}\,\text{eV}$$
this process will dominate over ICS on URB or CMB photons. The above threshold is not very strict as it depends on the 
URB density which is not a very well known quantity. The emitted gammas have a typical energy given by\cite{Batt_Sigl}
$$E_{\text{synch}} = 6.3\times 10^{11}\left(\frac{E}{10^{19}\,\text{eV}}\right )
\left ( \frac{B}{10^{-9}\,\text{G}}\right )\text{eV}.$$
Again low energy photon flux measurements will put constraints on the extragalactic fields and/or on the initial photon 
flux. 

Above threshold, the synchrotron radiation will damp the electron-positron pair energy 
extremely quickly. At 100~EeV in a $10^{-9}$~G magnetic field the attenuation length is of the order of 20~kpc. 
If one observes gammas above \E{20} they could not be high energy secondaries (e.g. from ICS) 
of an even higher energy photon converted into a pair. They must instead be primary ones. 
Consequently, their flux $j_\gamma(E)$ per unit area and unit solid 
angle at a given energy is directly related to the source distribution without any transport nor cosmological effects in between~:
$$ j_\gamma(E)
%~d\Omega = d\Omega\int_0^{l_\gamma(E)}\frac{\phi(E)}{4\pi r^2}r^2dr 
\sim \frac{1}{4\pi}l_\gamma(E)\phi(E)$$
%~d\Omega,$$
where $\phi(E)$ is the source emission density per unit time and energy interval and $l_\gamma(E)$ is the 
photon interaction length.  

Of course quantitative predictions of such effects is pending definite measurements of the galactic and
extragalactic magnetic fields. Although the magnetic fields of the galactic disc 
are now believed to be fairly well known 
this is not the case of the ones in the halo or extragalactic media. As mentioned in 
Section~\ref{Xfield}, several authors
advocate our bad knowledge of those fields in explaining the puzzling observational data and question both 
the typical value of $10^{-9}$~G and the coherence length 
of 1~Mpc usually assumed for the extragalactic fields.\cite{Sigl_Lemoine_Biermann}

\subsection{Conventional acceleration: Bottom-Up scenarios}\label{BottomUp}
\noindent
One essentially distinguishes two types of acceleration mechanisms:
\begin{itemize}
\item Direct, one-shot acceleration by very high electric fields. This occurs in or near 
very compact objects such as highly magnetized neutron stars or the accretion disks of black holes. 
However, this type of mechanism does not naturally provide a power-law spectrum. 

\item Diffusive, stochastic shock acceleration in magnetized plasma clouds which generally 
occurs in all systems where shock waves are present such as supernova remnants or radio galaxy hot spots. 
This statistical acceleration is known as the Fermi mechanism of first  (or second) order, depending on whether
the energy gain is proportional to the first (or second) power of $\beta$, the shock velocity.   
\end{itemize} 
Extensive reviews of acceleration mechanisms exist in the literature, e.g. on acceleration 
by neutron stars,\cite{Olinto2}  shock acceleration and propagation,\cite{Protheroe}  
non relativistic shocks,\cite{Drury}  and relativistic shocks.\cite{Kirk_Duffy}

\begin{figure}[!htb]

\begin{center}  
\epsfig{file=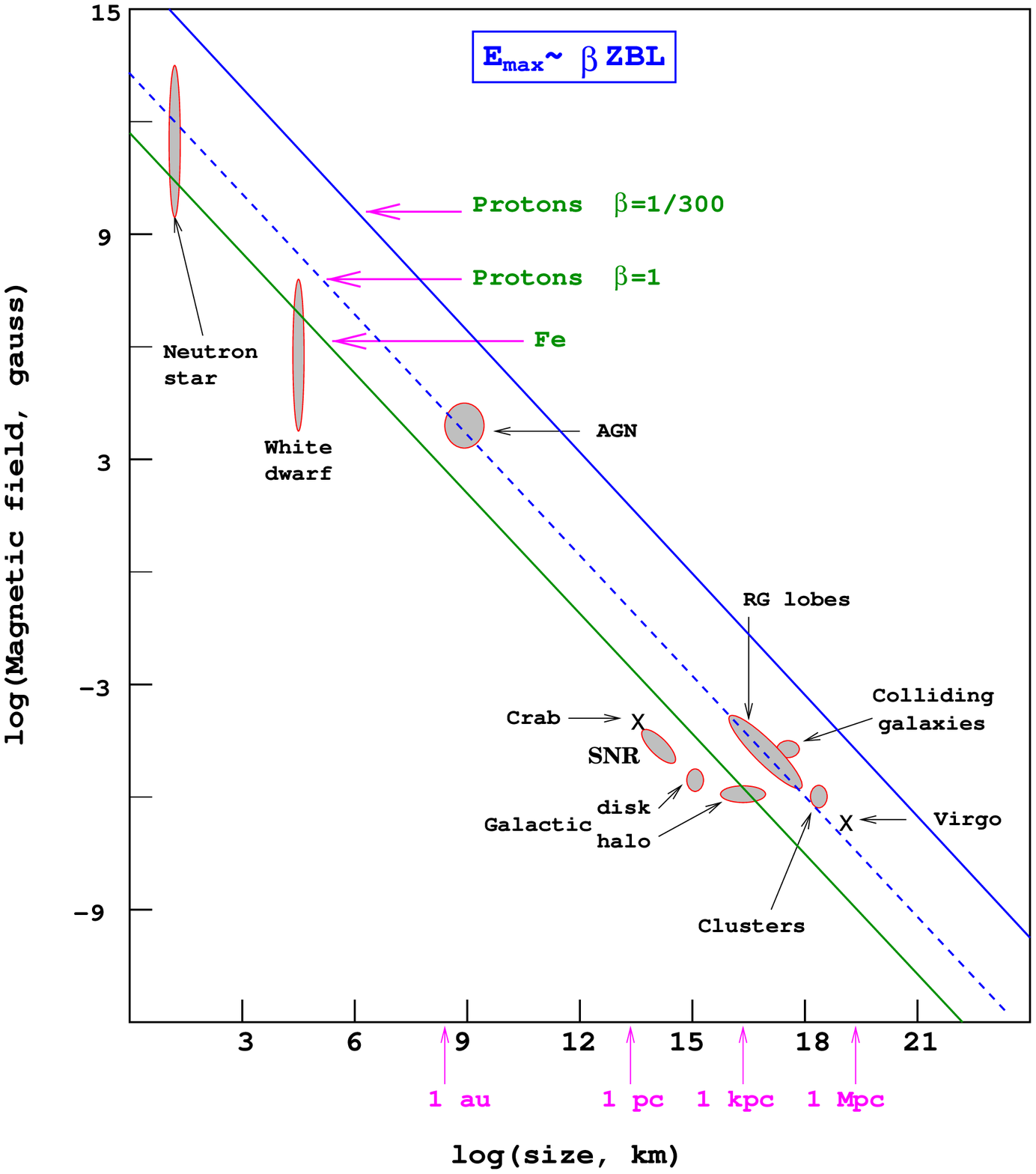, width = 8cm}
\end{center}

\fcaption{Size and magnetic field strength of possible acceleration sites. Objects below the diagonal 
lines cannot accelerate the corresponding elements (Iron with $\beta=1$ or protons $\beta=1$ and 
$\beta=1/300$) above \E{20}.\label{Hillas-Diagram}}
\end{figure}

Hillas has shown\cite{Hillas} that irrespective of the details of the acceleration 
mechanisms, the maximum energy of a particle of charge $Ze$ within a given site of size $R$ is: 
\begin{equation}
E_{\text{max}}\approx\beta Z\left(\frac{B}{1\,\mu \text{G}}\right )\left(\frac{R}{1\,\text{kpc}}\right)10^{18}~\text{eV}
\label{eq:Hillas}
\end{equation}
where $B$ is the magnetic field inside the acceleration volume and $\beta$ the velocity of the shock wave 
or the efficiency of the acceleration mechanism. This condition essentially states that the Larmor radius of the 
accelerated particle must be smaller than the size of the acceleration region,
%\fix
%It applies to both shock acceleration and ``one-shot'' 
%{\bf Xavier: mais pas \`a cause du rayon de Larmor pour les
%one-shot. \`a reformuler ?}}
%mechanisms (rotating neutron star) 
and is nicely represented in the Hillas diagram shown in 
Figure \ref{Hillas-Diagram}.   

\subsubsection{Candidate sites}\noindent
Inspecting the Hillas diagram one sees that only a few astrophysical sources satisfy the necessary, but not sufficient, 
condition given by Eq.~(\ref{eq:Hillas}). Some of them are reviewed e.g. by Biermann.\cite{Biermann} Let us 
just mention, among the possible candidates, pulsars, Active Galactic Nuclei 
(AGN) and  Fanaroff-Riley Class~II (FR-II) radio galaxies. 

\begin{quote}
\par {\it Pulsars}~\\
From a dimensional analysis, the electric field potential drop in a rotating magnetic pulsar is given by:
\begin{equation}
\Delta \Phi = \frac{B\times R^2}{\Delta T} 
\label{eq:pulsar}
\end{equation} 
One obtains $e\Delta \Phi=100$~EeV with $B=10^{9}\,$T, $\Delta T=10^{-3}\,$s and $R=10^4\,$m. 
However the high radiation density in the vicinity of the pulsar will produce  
$e^+e^-$ pairs from conversion in the intense magnetic field.\cite{Olinto2} 
These pairs will drift in opposite directions along the field lines and 
short circuit the potential drop down to values of about \E{13}. Moreover in the above dimensional analysis 
a perfect geometry is assumed. Actually, a more realistic geometry would introduce an 
additional factor $R/c\Delta T\sim 0.1$,  and further decrease the initial estimate. 
Finally, as will be described in the next section, synchrotron radiation losses in such 
compact systems become very important even for protons. 
\vspace{0.2cm}

{\it AGN cores and jets}~\\
Blast wave in AGN jets have typical sizes of a few percent of a parsec with magnetic fields of the 
order of 5 gauss.\cite{Zas} They could in principle  
lead to a maximum energy of a few tens of EeV. Similarily for AGN cores with a size of a few 
$10^{-5}$~pc and a field of order $10^3$~G one reaches a few tens of EeV.
However those maxima, already marginal, are unlikely 
to be achieved under realistic conditions. The very high radiation fields in and around the 
central engine of an AGN will interact with the accelerated protons 
producing pions and $e^+e^-$ pairs. Additional energy loss due to synchrotron radiation 
and Compton processes lead to a maximum energy of about \E{16}, way below 
the initial value.\cite{Batt_Sigl} To get around this problem, the 
acceleration site must be away from the active center and in a region with a lower radiation density such as in the terminal shock sites of the jets: a requirement possibly fulfilled by FR-II radio galaxies.
\vspace{0.2cm}

{\it FR-II radio galaxies}~\\
Radio-loud quasars are characterized by a very powerful central engine ejecting matter along 
thin extended jets. 
At the ends of those jets, the so-called hot spots, the relativistic shock wave is believed to be able to accelerate particles
up to ZeV energies. This estimate depends strongly on the value assumed for the spots' local
magnetic field, a very uncertain parameter. Nevertheless FR-II galaxies seem the best 
potential astrophysical source of EHECR.\cite{Biermann} Unfortunately, no nearby (less than 100~Mpc) 
object of this type is visible in the direction of the observed highest energy events. 
The closest FR-II source, actually in the direction of the Fly's Eye event at 320~EeV, is at about 
2.5~Gpc, way beyond the GZK distance cuts for nuclei, protons or photons.
\end{quote}
\subsubsection{Additional constraints}
\noindent
In addition to the constraint given by Eq.~(\ref{eq:Hillas}), candidate sites must also 
satisfy two additional conditions.
\begin{itemize}
\item The acceleration must occur on a reasonable time scale, e.g. the size of the acceleration 
region must be less than the interaction length of the accelerated particle. This is a 
relatively weak constraint since all the objects in the Hillas 
diagram have a size below 1 Mpc. However, in shock acceleration mechanisms the rate 
of energy loss on the CMB must be less than the rate of energy gain: 
\begin{equation}
-\frac{dE_{\text{loss}}}{dt} \propto \text{const}\times E < \frac{dE_{\text{gain}}}{dt} 
\end{equation}

\item The acceleration region must be large enough so that synchrotron losses are negligible compared to the 
energy given by Eq.~(\ref{eq:Hillas}). For shock acceleration the radiated synchrotron power must be below the rate of 
energy gain:
\begin{equation}
-\frac{dE_{\text{sync}}}{dt} \propto \frac{E^4}{R^2}\propto  B^2~E^2
\end{equation}
\end{itemize}

\begin{figure}[!htb]

\begin{center}  
\epsfig{file=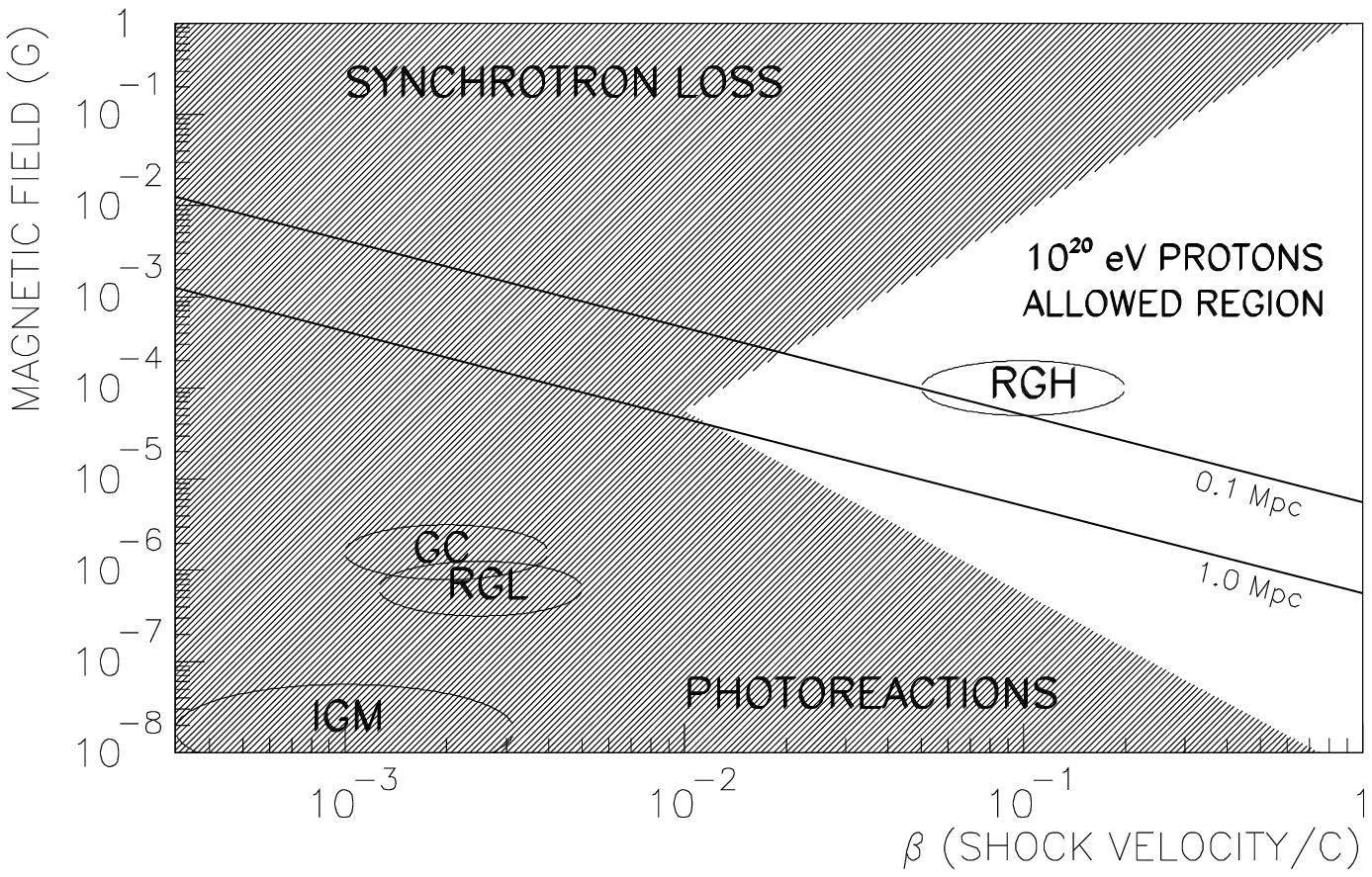, width = 8.5cm}
\end{center}

\fcaption{Magnetic field strength and shock velocity of possible sites. 
GC refers to Galactic Cluster (accretion shocks), IGM to Inter 
Galactic Medium, RGL to Radio Galaxy Lobes and RGH to Radio Galaxy Hot Spots
(a subclass of RGL).\cite{Auger}\label{adrf23}}
\end{figure}

Using, as the characteristic acceleration time, $T_A = R / \beta$ (where $\beta$ is 
the shock velocity) one finds a characteristic gain rate of :
\begin{equation}
\frac{dE_{\text{gain}}}{dt} \approx \frac{E_{\text{max}}}{T_A} \propto \beta^2~B.
\end{equation}
For a given $E_{\text{max}}$ (e.g. 100~EeV), these two constraints define two lines in the 
$\log{B}$, $\log{\beta}$ plane above and below which particles cannot be accelerated at 
the required energy. As can be seen on Figure~\ref{adrf23} the only
remaining candidates are the radio galaxy hot spots (RGH).

\vspace{0.5cm}
\par
\noindent
To conclude on the bottom-up scenario, let us mention a recent analysis from Farrar and 
Biermann.\cite{Farrar_Biermann} They have shown that, on cosmological scales, the 
correlation between the arrival direction of the five highest energy events and 
Compact Quasi Stellar Objects (CQSO's) which include radio-loud galaxies is 
unlikely to be accidental. They calculate the chance alignment probability to be $5\times 10^{-3}$. 
Only a new type of neutral particle could travel on distances over 1~Gpc without 
losing its energy on the CMB nor being deflected by extragalactic magnetic fields. 
More data at very high energy are needed to validate this result which would sign the existence 
of a new particle physics phenomenon. 

\newcommand{\X}{$X$}
\subsection{``Exotic" sources: Top-Down scenario}\label{TopDown}
\noindent
One way to overcome the many problems related to the acceleration of EHECR, 
their flux, the visibility of their sources and so on, is to 
introduce a new unstable or meta-stable super-massive particle, currently called the \X-particle. 
The decay of the \X-particle produces, among other things, quarks and leptons. 
The quarks hadronize, producing jets of hadrons which, together with the decay products of the 
unstable leptons, result in a large cascade of energetic photons, neutrinos and light leptons with a small 
fraction of protons and neutrons, part of which become the EHECR. 

For this scenario to be observable three conditions must be met:
\begin{itemize}
\item The decay must have occurred recently since the decay products must have traveled less than about 100~Mpc because of the attenuation processes discussed above.
\item The mass of this new particle must be well above the observed highest energy (100~EeV range), 
a hypothesis well satisfied by Grand Unification Theories (GUT) whose scale is around $10^{24}$-\E{25}.
\item The ratio of the volume density of this particle to its decay time must be 
compatible with the observed flux of EHECR. 
\end{itemize}
\noindent
The \X-particles may be produced by way of two distinct mechanisms:
\begin{itemize}
\item  Radiation, interaction or collapse of Topological Defects (TD), producing \X-particles that
 decay instantly. In those models the TD are leftovers
from the GUT symmetry breaking phase transition in the very early universe.
However very little is known on the phase transition itself and on the TD density that survives a
possible inflationary phase, and quantitative predictions are usually quite difficult to rely on.

\item  Super-massive metastable relic particles from some primordial quantum field, produced after
the now commonly accepted inflationary stage of our Universe. Lifetime of those relics should be of the order
of the age of the universe and must be guaranteed by some almost conserved protecting symmetry.
It is worth noting that in some of those scenarios the relic particles
may also act as non-thermal Dark Matter.
\end{itemize}

In the first case the \X-particles instantly decay and the flux of EHECR is related to their production rate
given by the density of TD and their radiation, collapse or interaction rate, 
while in the second case the flux is driven by the ratio of the density of the relics over their 
lifetime. In the following
the  terms  {\it ``production or decay rate''} will refer to these two situations.
Before discussing the exact nature of the \X-particles we shall briefly review the main characteristics
of the decay chain and the expected flux of the energetic outgoing particles.

\subsubsection{\X decay and secondary fluxes}
At GUT energies and if they exist,  squark and 
sleptons are believed to behave like their corresponding super-symmetric partners 
so that the gross characteristics of the cascade may be inferred from the 
known evolution of the quarks and leptons. Of course the internal mechanisms of the decay and 
the detailed dynamics of the first secondaries do depend on the exact nature 
of the particles but the bulk flow of outgoing particles is most certainly independent of such details.\cite{Batt_Sigl} 

\begin{figure}[!htb]

\begin{center}  
\epsfig{file=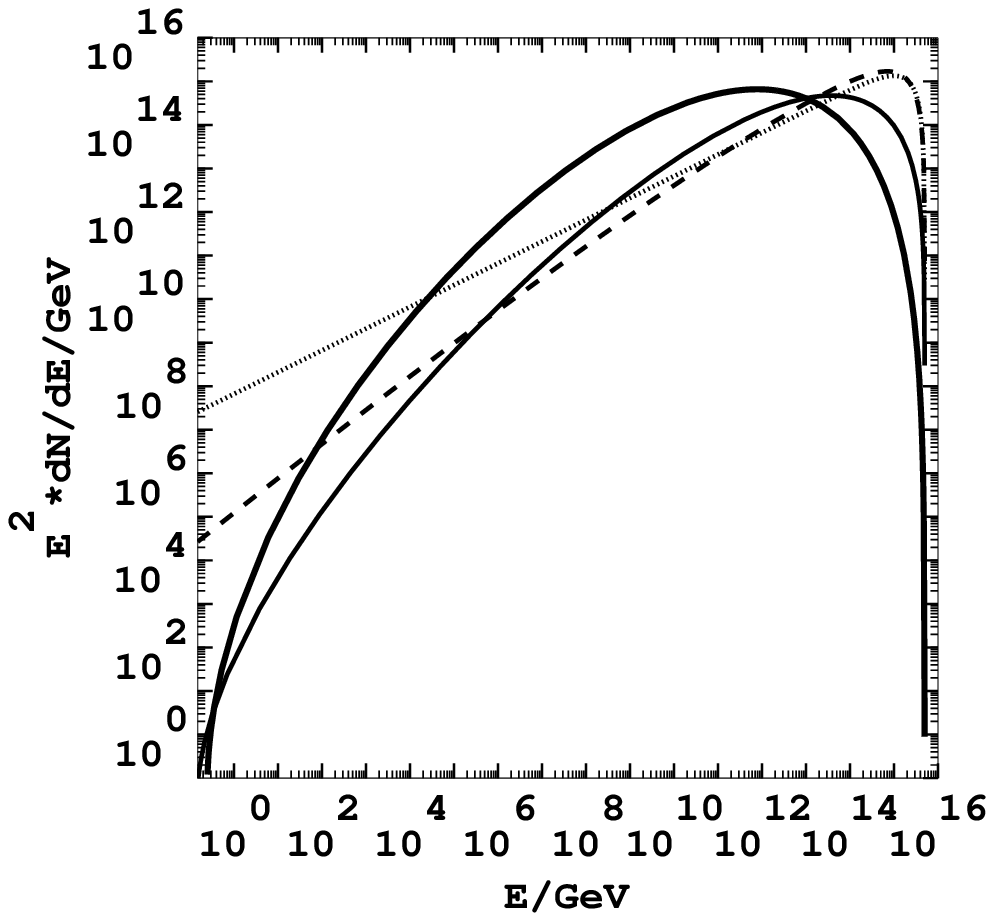, width = 8cm}
\end{center}

\fcaption{Fragmentation function in the Modified Leading Log Approximation for a total 
jet energy of $5\times$\E{24} with SUSY (thick solid line peaking at $10^{12}$GeV) and 
without SUSY (thin solid Line) as calculated by Bhattacharjee and Sigl.\cite{Batt_Sigl} 
Other lines (dashed and dotted) are older approximated expressions.  
%Figure taken from\cite{Batt_Sigl}.
\label{Sigl23}}
\end{figure}

A common picture for the  hadronisation of the decay products follows three steps. 
At the high energy end, the perturbative QCD expansion provides a good description 
of the hard processes driving the dynamics of the parton cascade. At a cutoff energy 
of about 1~GeV soft processes become dominant and partons are glued together to form 
color singlets which will in turn decay into known hadrons. The LUND\cite{Lund}
string fragmentation model provides a description for the second and last phases while 
a model like the Local Parton-Hadron Duality directly relates the parton density in the parton cascade to 
the final hadron density.\cite{Lphd} Nevertheless and despite the fact that up to 40\% 
of the initial energy may turn into LSP, the cascade produces a rather 
hard\footnote{For a power law spectrum of exponent $\alpha<2$ the total energy ($\propto E^{2-\alpha}$)
is dominated by the high energy end of the integral, i.e. a few very energetic particles, thus a hard spectrum, while for $\alpha>2$ the energy is carried by the very large number of low energy particles, i.e. a soft spectrum.}
~hadron spectrum adequately described by:
$$ \frac{dN_h}{dE} \propto E^{-\alpha}\mbox{~~with~~} 1<\alpha<2$$
in the range $E/m_X \ll 1$, where $m_X$ is the \X-particle mass. At the high energy end a cutoff occurs 
at a value depending on the \X-particle 
mass and on the eventual existence of new physics such as Super Symmetry (SUSY), which would displace 
the maximum of the hadron spectrum to a lower energy (see Figure~\ref{Sigl23}).

Indeed, Super Symmetry is not the only possible candidate theory for new physics beyond the standard 
model: other (yet unknown) models may appear as possible alternatives in the future. 
However, in all cases, secondaries from Top-Down mechanisms should
manifest themselves as a change of slope in the EHECR spectrum, above 10~EeV and over a range which 
will reflect the (new) physics at play. 

In all conceivable Top-Down scenarios, photons and neutrinos dominate at the end of the hadronic cascade. 
This is \emph{the} important distinction from the conventional acceleration mechanisms.
The spectra of  photons and neutrinos can be derived from the charged 
and neutral pion densities in the jets as:
$$\Phi_{\gamma}^{\pi^0}(E,t) \simeq 2 \int_E^{E_{\text{jet}}}\Phi_{\pi^0}(\varepsilon,t)d\varepsilon/\varepsilon$$
$$\Phi_{\nu}^{\pi^\pm}(E,t) \simeq 2.34 \int_{2.34E}^{E_{\text{jet}}}\Phi_{\pi^\pm}(\varepsilon,t)d\varepsilon
/\varepsilon$$
where $E_{\text{jet}}$ is the total energy of the jet (or equivalently the initial parton energy). Since 
$\Phi_{\pi^\pm}(\varepsilon,t)\simeq 2\Phi_{\pi^0}(\varepsilon,t)$, photons and neutrinos should have very 
similar spectra. These injection spectra must then be convoluted with the transport phenomena to obtain 
the corresponding flux on Earth. As was mentioned in Section \ref{photon} the photon 
transport equation strongly depends on its energy and on the badly known Universal 
Radio Background and extragalactic magnetic fields. 

\subsubsection{$X$ production or decay rates: a lower limit}
\noindent
The production or decay rates of the \X-particles are very model dependent and no firm  
prediction on the expected flux of EHECR can be made. However, in their review, Bhattacharjee 
and Sigl evaluate with a simple model the rate needed to explain the observed EHECR 
fluxes. Assuming that photons dominate at the source and on Earth and that they follow 
a power law spectrum of index $\alpha$; assuming also that the 
initial \X~decay secondaries are quarks and leptons in equal numbers, they calculate 
a lower limit on the production rate given by (for $\alpha=1.5$):
\begin{equation}
\dot{n}_X \geq 10^{-46} \left( \frac{\mbox{10\,Mpc}}{l_E(E_\gamma)}\right )\left( \frac{E^2j_\gamma(E
)}{{\cal F}_\oplus}\right )\sqrt{\frac{m_X}{10^{16}\,\text{GeV}}} ~\text{cm}^{-3}\text{s}^{-1}
\label{prod_lim}
\end{equation}
Here ${\cal F}_\oplus \approx 1$~eV$\,$cm$^{-2}$s$^{-1}$sr$^{-1}$ is the observed 
energy flow of EHECR at 100~EeV and $l_E(E_\gamma)$ the photon attenuation length. 
Additional normalization factors of order unity have not been reproduced here. 
In other words, for TD or relics to explain the observed EHECR
flux at 100~EeV and assuming an \X~mass of $10^{16}$~GeV their production or decay rates 
must be larger than $10^{-46} \text{cm}^{-3}\text{s}^{-1}$. This is of course only 
an order of magnitude calculation which may be modified by 
the decay dynamics and the distribution of the \X-particles, but can be used 
as a reasonable benchmark of the necessary rates.
 
\subsubsection{More about \X-particles}
\noindent
{\it Topological defects}\\
The very wide variety of topological defect models together with their large number of parameters makes them 
difficult to review in detail. Many authors have addressed this field. Among them let us mention 
Vilenkin and Shellard\cite{Shell} and Vachaspati\cite{Vach1,Vach2} for a review on TD formation 
and interaction, and Bhattacharjee,\cite{Bhatt} Bhattacharjee and Sigl\cite{Batt_Sigl} 
and Berezinky, Blasi and Vilenkin\cite{Berez} for a review on experimental signatures in 
the framework of the EHECR. 

According to the current picture on the evolution of the Universe, several symmetry 
breaking phase transitions such as $GUT\Longrightarrow H~...\Longrightarrow SU(3)\times 
SU(2)\times U(1)$ occurred during the cooling. For those ``spontaneous'' symmetry 
breakings to occur, some scalar field (called the Higgs field) must acquire a non vanishing 
expectation value in the new vacuum (ground) state. Quanta associated to those fields have 
energies of the order of the symmetry breaking scale, e.g. $10^{15}-10^{16}$~GeV for the 
Grand Unification scale. Such values are indeed perfectly in the range of interest 
for the above mentioned \X-particles.

During the phase transition process, non causal regions may evolve towards different 
states in such a way that at the different domain borders, the Higgs field keeps a null expectation value. 
Energy is then trapped in a TD whose properties depend on the topology of the manifold where 
the Higgs potential reaches its minimum (the vacuum manifold topology). 

Possible TDs are classified according to their dimensions: magnetic mo\-no\-poles (0-dimensional, point-like); 
cosmic strings (1-dimensional); a sub-variety of the previous which carries current and is supra-conducting; 
domain walls (2-dimensional); textures (3-dimensional). Among those, only monopoles 
and cosmic strings are of interest as possible EHECR sources: textures do not 
trap energy while domain walls, if they existed, would over-close the Universe.\cite{Zel}

In GUT theories, magnetic monopoles always exist because the reduced symmetry group contains at least the 
electromagnetic $U(1)$ invariance. In fact it is the predicted over abundance of magnetic monopoles 
in our present universe that led Guth\cite{Guth} to come up with the now well adopted idea of an 
inflationary universe. Strings on the other hand are the only defects that can be relevant for structure 
formation. It is possible, from the scaling property of the string network,
to relate the string formation scale $\eta$ to the mass fluctuations in the Universe.
Using the large scale mass fluctuation value of $\delta M/M \sim 1$ this gives $\eta\simeq10^{16}$~GeV and similar 
conclusions are drawn if one uses the COBE results on CMB anisotropies.\cite{Brand} 
It is striking to see 
that if strings were to play a role in large scale structure formation, hence making the Hot Dark Matter 
scenario viable, the energy scale at which this would be possible also corresponds to 
that relevant for EHECR production.

When two strings intercommute, the energy release sometimes leads to the 
production of small loops that will release more energy when they collapse. These are, among other mechanisms,
fundamental dissipation processes that prevent the string network from dominating the energy density in the Universe. 
For monopoles, it is the annihilation of monopolonia (monopole-antimonopole bound 
states)\cite{Hill,Schramm_Hill}  that releases energy.\footnote{In fact monopolonia  
are too short lived but monopole-anti-monopole pairs connected by a string have appropriate lifetime. This happens when the $U(1)$ 
symmetry  is further broken into $Z_2$.}~~In each case part of the released 
energy is in the form of \X-particles.    

Strings and monopoles come in various forms according to the scale at which TDs are formed
and to the vacuum topology. They may even coexist. Nevertheless, the \X-particle production rate may, 
on dimensional grounds, be parametrized in a very general way.\cite{Bhatt_Hill_Schramm} 
Introducing the Hubble time $t$, the production rate can be written as:
\begin{equation}
\dot{n}_X(t) = \frac{Q_0}{m_X}\left (\frac{t}{t_0}\right )^{-4+p}
\label{X_prod}
\end{equation}
where $Q_0\equiv \dot{n}_X(t_0)\,m_X$ is the energy injection rate at $t=t_0$ (the present epoch). The parameter $p$ 
depends on the exact TD model. In most cases (intercommuting strings, collapsing loops as well as 
monopolonium annihilation) $p=1$ but superconducting string models can have $p\leq 0$ while decaying 
vortons\footnote{Superconducting string loops stabilized by the angular momentum of the charge 
carriers.}~~give $p=2$.  

One can compare the integrated energy release of Eq.~(\ref{X_prod}) in the form of 
low energy (10~MeV - 100~GeV) photons resulting from the cascading of the electromagnetic component 
of the $X$-particle decay into the diffuse extragalactic gamma ray 
background, $w_{\text{em}}\sim 10^{-6}$~eV~cm$^{-3}$~s$^{-1}$, as measured by EGRET. 
Assuming as in Ref.\cite{Berez} that half of the energy release goes into the electromagnetic 
component, one obtains~:
\begin{equation}
w_{\text{em}} = \frac{Q_0}{2}\int_{t_{\text{min}}}^{t_0} \left( \frac{t}{t_0}\right )^{-4+p} \frac{dt}{(1+z)^4}.
\label{wem_prod}
\end{equation}
where $(1+z) = (t_0/t)^{2/3}$ in a matter dominated Universe.
For $\alpha\equiv p -1/3 > 0$ evolutionary effects are negligible and Eq.~(\ref{wem_prod}) simply leads to 
$$
w_{\text{em}} \simeq \frac{Q_0}{2}\frac{t_0}{\alpha}
\label{wem_1}
$$
or, using the EGRET limit and $t_0\simeq 2\times 10^{17}h^{-1}$s, to:
$$
\dot{n}_X(t_0) \leq \alpha \frac{10^{-48}h~\text{cm}^{-3}\text{s}^{-1}}{m_X/10^{16}\,\text{GeV}}
$$
a limit hardly compatible with the order of magnitude given by 
Eq.~(\ref{prod_lim}). However, more information about the EHECR fluxes, 
the diffuse gamma ray background and extragalactic magnetic fields are needed to confirm this trend.

In the models where $\alpha<0$ evolutionary effects can become important. In fact it is the lower bound of the 
integral of Eq.~(\ref{wem_prod}) that would dominate. Using as a lower bound the decoupling time
$t_{\text{dec}}/t_0\sim 10^{-5}$ one gets:
$$
w_{\text{em}} \simeq \frac{Q_0}{2}\left (\frac{t_{\text{dec}}}{t_0}\right )^{-\alpha}\frac{t_0}{|\alpha|}
\label{wem_2}
$$
or,  
$$
\dot{n}_X(t_0) \leq |\alpha|\frac{10^{-48-5\alpha}\text{cm}^{-3}\text{s}^{-1}}{m_X/10^{16}\,\text{GeV}}.
$$
which,  for $p\leq 0$ is perfectly compatible with Eq.~(\ref{prod_lim}). 
However the large density of gamma rays 
released in the early Universe impacts on the $^4$He production and on the uniformity of the CMB making this 
kind of models currently unfavored in the context of EHECR.

\vspace*{0.5cm}
\noindent
{\it Supermassive relics}\\
Supermassive relic particles may be another possible source of EHECR.\cite{Berez1} Their mass should be 
larger than $10^{12}$~GeV and their lifetime of the order of the age of the Universe since 
these relics must decay now (close by) 
in order to explain the EHECR flux. Unlike strings and monopoles, but like monopolonia, 
relics aggregate under the effect of gravity like ordinary matter and act as a (non thermal) 
cold dark matter component. The distribution of such relics should consequently be biased 
towards galaxies and galaxy clusters. A high statistics study of the EHECR arrival distributions 
will be a very powerful tool to distinguish between aggregating and non-aggregating Top-Down sources. 

If one neglects the cosmological effects, a reasonable assumption on the decay rate would 
simply be, since the decay should occur over the last 100 Mpc/c:
$$
\dot{n}_X = \frac{n_X}{\tau}
$$
where $\tau$ is the relic's lifetime and where the relic density $n_X$ may be given in 
terms of the critical density of the Universe $\rho_c$ as:
$$
n_X = \frac{\rho_c(\Omega_X h^2)}{m_X} = 10^{-17} (\Omega_X h^2) \left (\frac{m_X}{10^{12}~\text{GeV}}\right )^{-1}
$$   
From which, with the constraint given by Eq.~(\ref{prod_lim}) and using $m_X=10^{12}$~GeV, one obtains a 
lifetime of the order of 
$10^{21}(\Omega_X h^2)$ years. To obtain such a value, orders of magnitude larger than 
the age of the Universe, one needs a  symmetry (such as $R$-parity) to be very weakly 
broken (wormhole effect, instanton induced decays) unless the  fractional abundance 
$\Omega_X$ represents only a tiny part ($\sim 10^{-11}$) of the density of the Universe, in which case the 
 production mechanism of relics must be extraordinarily inefficient.

\subsection{Conclusions}
\noindent
The cosmic rays' chemical composition, the shape of their energy spectrum and the distribution of their directions of arrival will prove to be powerful 
tools to distinguish between the different acceleration or decay scenarios. 

If the EHECR are conventional hadrons accelerated by Bottom-Up mechanisms, they should point 
back to their sources, with a quite specific distribution in the sky and a spectrum clearly 
showing the GZK cutoff.  If, on the other hand,  the accelerated particles 
are not conventional, they should in any case be charged particles (otherwise they can only 
be secondary collision products) weakly interacting with the CMB but strongly with the atmosphere. 

For Top-Down mechanisms and above the ZeV, one should observe a flux of photons 
(and neutrinos) as the photon absorption length increases (up to several Gpc). Below 
100~EeV the spectrum shape will depend on the relative values of, the characteristic distance between 
TD interactions or relic particle decays and Earth ($D$), the proton
attenuation length ($R_p$), and the photon absorption 
length ($L_\gamma$).  
Following the description of Ref.\cite{Berez} the following situations can be disentangled:
\begin{itemize}
\item $R_p<D$: a very low flux with an exponential cutoff. If the sources are nearby,
the observed distribution will be strongly anisotropic.

\item$L_\gamma<D<R_p$: the protons dominate and the GZK cutoff is visible. As energy 
increases, the direction of arrival distribution
 should become more and more anisotropic as photons no longer get absorbed.

\item $D<L_\gamma$: a very strong flux in the direction of the sources; photons dominate.

\item $D\ll L_\gamma$: the GZK cutoff is visible and protons dominate as long as $R_p(E)$ is much 
larger than $L_\gamma(E)$. Photons dominate  above a few ZeV. The arrival distribution is isotropic 
at all energies.
\end{itemize} 
For relic particles and TDs like vortons and monopolonia, because of the accumulation in the galactic 
halo, photons will dominate the flux over the extragalactic contribution. Some anisotropy 
should be visible due to the earth's  slightly eccentric position in the halo. The spectrum will not 
show any GZK cutoff and the EGRET constraints on the injection 
rate will not apply as the emitted photons have no time to cascade over the short distances.

Finally, if nuclei can possibly be EHECR candidates in Bottom-Up scenarios, they are completely 
excluded in the Top-Down cascades.

\section{Next} 
\noindent 
The current EHECR studies essentially rely on the AGASA detector, described
in section~\ref{agasa_det}.
Since 1990, the date of its start, it has reported about 40 events with
energies above 40~EeV out of which 7 exceed 100~EeV. Well adapted for the exploration of the spectrum between 1 and 50~EeV, its
major role for the physics beyond the GZK cutoff is the confirmation that
something new is happening there. A systematic exploration of this region 
hopefully bringing an answer to the question ``What and How?" will come from the
detectors and projects that we will describe here. To compare the performances
of the future detectors or projects, we propose to use, as a reference, the
flux rate formula given in Eq.~\ref{eq:flux}.

\subsection{The present: HiRes\label{hires}}
\noindent
The ``High Resolution Fly's Eye" or HiRes\cite{hires} is an improved version of
the original Fly's Eye detector (section ~\ref{flyseye}). The present
version of the detector design consists of two ``Eyes" separated by 13~km,
situated at a military site in Dugway (Utah, USA). The basic fluorescence
telescope of HiRes is a mirror equipped with a camera of 256 phototubes, each
phototube (pixel) watching an angular region of space of
$1^{\circ}~\times~1^{\circ}$. Therefore, each telescope has a field of
view of $16^{\circ}~\times~16^{\circ}$, and a complete eye is expected to cover a field of $30^{\circ}$ in
elevation and $360^{\circ}$ in azimuth with 44 telescopes. The collaboration
expects to have one full and one half eyes 
by the end of 1999. The fluorescence light emitted by the EAS can be seen
by such telescopes at distances up to 20~km or more depending on their energy. The aperture of the detector
has to be convoluted with its duty-cycle (10\%),
and is energy-dependent (far showers can be reconstructed only if they are in
the higher energy range). Once fully equipped its aperture should be of 
350~km$^2\,$sr at 10~EeV and about 1000~km$^2\,$sr at 100~EeV. This gives a detection
rate (based on the empirical formula, see above) of about 10 events/year above
100~EeV. The HiRes detector's optimal operating range is therefore the decade
between 5 and 50~EeV.

It is interesting to note that the HiRes prototype is overlooking the Chicago
AirShower Array (CASA) also installed at Dugway, thus allowing the detection of
a few events in the ``hybrid" mode.

\subsection{Starting: Auger Observatory}
\noindent
The Auger Observatory\footnote{Named after the french physicist Pierre Auger, see
Section~\ref{Auger}.}~ is the only existing or projected
detector whose design is based on the ``hybrid" detection mode (EAS
simultaneously observed by a ground array and a fluorescence
detector).\cite{Auger}

The Observatory whose construction starts during the fall of 1999, once completed, 
will be covering two sites, respectively
in the southern (Pampa Amarilla, province of Mendoza, Argentina) and  northern
(Millard County, Utah, USA) hemispheres. The southern hemisphere
detector is especially interesting since very few detectors took data in the
past in this part of the world from where the direction of the center of our
galaxy is visible.

The size of the ground array is adapted to its physics aims: explore the
spectrum around and above the GZK cutoff. Therefore the surface of each site
was chosen to be of 3000~km$^2$, so as to provide a statistics of a few tens of
expected events per year above 100~EeV. The detector is designed to be fully
efficient for showers with energies of 10~EeV and above, with a duty-cycle of
100\%. This will make the link with the part of the energy spectrum well
explored by presently operating or upcoming detectors, AGASA and HiRes. The
energy threshold defines the spacing of the detector ``stations": with a
spacing of 1.5~km between the stations, a 10~EeV vertical shower will hit on
average 6 stations which is enough to fully reconstruct the EAS.
With such a spacing on a regular grid, the total number of stations is about
1600. Each station is a cylindrical tank (the same basic design as the Haverah
Park array, but with much more sophisticated electronics), of 10~m$^2$ 
surface and 1.2~m height. The tanks are filled with filtered
water in which the secondary particles from the EAS produce light by Cerenkov
radiation. The light is reflected and diffused by an internal coating and
detected by three phototubes installed on the top. Flash-ADCs
cycling at a rate of 40~MHz record the pulse heights as a function of time.
With such a system, the separation of the muons from the electromagnetic
component of the shower becomes reasonably good.

Because of the size of the array, the stations have to work in a stand-alone
mode: they are powered by solar panels and batteries, and the communication with
the central station where the data-taking system is installed is done by
telecommunication techniques.

The giant array is completed by an optical device detecting the fluorescence
light emitted by the nitrogen molecules of the atmosphere excited by the
charged particles produced in the EAS. The fluorescence telescopes use pixels
(phototubes) with a field of view of $1.5^{\circ}$. A telescope is  a
camera with 400 pixels installed at the focal plane of a mirror whose size is
$3.5\times3.5$~m$^2$. Each telescope sees an angle of about $30\times30$
degrees. On the southern site, three eyes (7 telescopes each) will be installed
at the periphery  of the array and one (12 telescopes) in the middle, in order for
the whole array to be  visible by at least one of the telescopes. On the northern site,
the optical component \emph{could} be the HiRes we described in the previous
section or the Telescope Array which we present in the next one.

In the hybrid mode (10\% of the events), the detector is expected to have on
average 10\% energy resolution and an angular precision of about $0.3^{\circ}$.
For the array alone those numbers become 20\% and less than $1^{\circ}$.

The statistics attainable with such a detector is dependent on what exactly
will happen above the GZK cutoff.  However, with its total aperture of
14000~km$^2$sr (both sites), the Auger Observatory should detect of the order of
6000 events above 10~EeV and 60 above 100~EeV per year.

\subsection{The future: the Telescope Array}
\noindent
The Telescope Array\cite{ta} (TA) is a project of a purely optical
detector 
which, in its initial design, aims to be used
as a gamma-astronomy device as well as a fluorescence detector. Its basic principles are identical to the HiRes
technique. The individual telescopes are made of 3 meter diameter segmented
mirrors with a camera of 256 pixels. Each pixel (hexagonal phototubes) sees a
patch of sky in a solid angle of $1^{\circ}\times 1^{\circ}$ (hence a
field-of-view of $16^{\circ}\times 16^{\circ}$ per telescope). One TA station
is made of 42 telescopes, thus viewing the sky over $2\pi$ in azimuth and 
$30^{\circ}$ in elevation. If the project is approved and completed, the TA
collaboration intends to install on the Utah site 8 stations in a snake-shaped
pattern which would extend from the present position of the HiRes detector in
Dugway to the future northern site of the Auger Observatory in Millard County
(about 100~km south of Dugway). Two of the stations would then become the
optical component of the Auger array to detect showers in the hybrid mode. The
total aperture of the array should be about 5 times the HiRes acceptance,
depending on the spacing of the stations and the performance of the pixels.

\subsection{The far future: Airwatch/OWL}
\noindent
One cannot exclude the possibility that the EHECR spectrum extends to even
higher energies than those for which the three above projects were designed.
Such a hypothesis would become likely if the production mechanism is one of the
top-down models, in which case the spectrum could go 
well above 1~ZeV. At the same time, the fluxes at these energies would be
still smaller than those detectable by Auger. Is there any means of observing
with reasonable statistics events in this energy range?

It is easy to see that the ground array method can hardly be envisaged if one,
or more, order of magnitude increase in aperture is needed. The aperture is
proportional to the surface of the array, and for practical (and to some
extent, economical) reasons the present surface of the Auger arrays is close to
what can be considered as an upper limit. As for the fluorescence
technique, there are two ways of increasing the aperture:
increase the number of telescopes and/or increase the sensitivity of the
pixels, i.e. the visible depth of field. The former solution is the one envisaged
by the Telescope Array collaboration. Here again the cost increases linearly
with the aperture. The latter solution is the basic idea of the Airwatch/OWL
projects. 

The physics aims of both projects is the same as that of the detectors outlined
in the preceding sections: solve the puzzle of the origin of the highest energy
cosmic rays. The technical solution proposed is similar to that of the Fly's
Eye: observe the air-shower's fluorescence light, but from \emph{above} (i.e.
by a detector installed in a dedicated or general-purpose satellite) and from a
distance of 500~km instead of the 20~km field of view for the ground-based
version.

The idea of observing the EAS from space originated from converging ideas of
J.Linslay at the end of the seventies and a proposal made by Y.Takahashi during
the mid-nineties for a Maximum-energy Air Shower observing Satellite (MASS).
The project was then promoted in parallel in Europe (mainly Italy) under the
name of Airwatch,\cite{airwatch} and in the USA (with the support of NASA and
participation of Japanese groups) in its OWL\cite{owl} (Orbiting Wide-angle
Light collectors) version. The basic idea is to use about $10^6$ highly
sensitive pixels at the focal plane of a wide angle ($30^{\circ}$ cone) optical
system (Fresnel lenses and reflectors). Several options are being studied for
the pixels. The expected performances would be: a resolution of 1~km from a
distance of 500~km, 1 photoelectron sensitivity of the pixels, very good
background rejection (city lights, atmospheric phenomena) and, finally, an
effective aperture of $10^5$~km$^2$sr, increasing that of the Auger Observatory
by a factor of 10. A good reconstruction of the showers, as with the
ground based fluorescence detectors, needs a stereo view, therefore two
satellites. The project still needs a large  R\&D effort. Adding to it the
funding and construction of the detector, it is reasonable to assume that the
it will not be operational before ten years or so.

\section{Conclusions}
\noindent
The EHECR were a puzzle when they were first observed, more than 30 years ago. 
They still are. In our review we tried to show that this statement is true to
a  large extent. This does not mean that theorists find no explanation as to
their  existence, but rather that their models are either not fully convincing
(in the  sense that they do not explain the full set of observed data in the
field) or  based on exotic, and as such still-to-be-confirmed, theories. 

The past experiments which explored this field could do hardly better than 
con\-vince us of the existence of the EHECR above the GZK cutoff. Statistics 
which should make us able to locate the sources, reconstruct the shape of  the
CR spectrum above the cutoff and study the CR chemical composition will  soon
be provided by the ongoing (HiRes, AGASA) and oncoming (Auger,
Telescope-Array,  OWL/Airwatch) experiments.\\
 
We shall leave the last (but hopefully provisional) word to
L.Celnikier\cite{Celnikier} who, in a review article written a few years ago, concluded: ``Contemporary astrophysics is faced by a number of acute  
problems. One of them concerns dark matter, which one might (perhaps  
mischievously) qualify as the study of particles which \emph{should}  
exist\ldots  \   but until farther notice,  don't. Ultra high energy cosmic rays constitute the inverse problem:   
particles which \emph{do} exist\ldots  \   but perhaps shouldn't.'' To our best knowledge, the statement is still actual both ways.

\nonumsection{Acknowledgements}  
\noindent 
The authors have made, for this article, extensive use of the work produced over 
the past three years in the framework of the Pierre Auger project.
We are grateful to many of the members of this collaboration and, indeed,
specifically  to the project spokespersons James W.Cronin and Alan A.Watson
for  their availability for so many stimulating discussions. Also many thanks
to P.Billoir and P.Sommers  for their useful and perceptive comments, to
P.Astier and E.Berman for their careful reading of the manuscript.

\nonumsection{References}
\noindent
The references given as astro-ph/xxxxxxx or hep-ph/xxxxxxx are articles
available from the Web electronic preprint archive at the URL\\ {\tt http://xxx.lanl.gov/}


\begin{thebibliography}{000}
\bibitem{Hemas} Web site: {\tt osf1.lngs.infn.it/\lower
3pt\hbox{\string~}carboni/Hemas/doc/}
\end{thebibliography}
\end{document}